\documentclass[12pt,preprint]{aastex}

\usepackage{graphics,epsfig}

\bibpunct{(}{)}{;}{a}{}{;}

\def\za{$\zeta$~Aur}
\def\fract#1/#2{\hskip1pt\raise.6ex\hbox{\the\scriptfont0 #1}\hskip-1pt
\hskip-1pt \lower.3ex\hbox{\the\scriptfont0 #2}}
\def\kms{km\,s$^{\rm -1}$}
\newcommand{\be}{\begin{equation}}
\newcommand{\ee}{\end{equation}}

\def\longref#1 {\par{\hangindent=20pt \hangafter=1 #1 \par}}

\shorttitle{Atmospheric Heating and Wind Acceleration}

\shortauthors{Airapetian \& Cuntz}

\begin{document}


\title{Atmospheric Heating and Wind Acceleration: \\
Results for Cool Evolved Stars based on Proposed Processes}

\author{Vladimir S. Airapetian}
\affil{NASA / Goddard Space Flight Center,}
\affil{8800 Geenbelt Rd., Greenbelt, Maryland 20771, USA}
\email{vladimir.airapetian@nasa.gov}
\and
\author{Manfred Cuntz}
\affil{Department of Physics}
\affil{University of Texas at Arlington, Arlington, Texas 76019-0059, USA}
\email{cuntz@uta.edu}

\begin{abstract}
A chromosphere is a universal attribute of stars of spectral type
  later than $\sim$F5.  Evolved (K and M) giants and supergiants (including the
  \za~binaries) show extended and highly turbulent chromospheres, which develop
  into slow massive winds.  The associated continuous mass loss has a
  significant impact on stellar evolution, and thence on the chemical evolution
  of galaxies.  Yet despite the fundamental importance of those winds in
  astrophysics, the question of their origin(s) remains unsolved.  What sources
  heat a chromosphere?  What is the role of the chromosphere in the formation
  of stellar winds?  This article provides a review of the observational
  requirements and theoretical approaches for modelling chromospheric heating
  and the acceleration of winds in single cool, evolved stars and in eclipsing
  binary stars, including physical models that have recently been proposed.  It
  describes the successes that have been achieved so far by invoking acoustic
  and MHD waves to provide a physical description of plasma heating and wind
  acceleration, and discusses the challenges that still remain.
\end{abstract}



\section{Introduction}

Stars of spectral types later than about F5 (including the Sun) possess
convective zones, chromospheres and winds. The strong convective motions in
giant and supergiant stars---thus also encompassing the primaries of
\za~systems---carry processed material, produced by nuclear-burning reactions,
from the stellar interior to the atmosphere, where massive winds originate, and
from there it is injected into the interstellar medium. Continuous mass loss
therefore has a significant impact upon stellar evolutionary patterns,
especially in the late phases of a star's life, on the chemical evolution of
galaxies (including the mass and energy budgets of the interstellar medium),
and even on the long-term evolution of exoplanetary atmospheres.  Mass loss
from stars can also affect planetary habitability through the dynamical and
collisional evolution of planetesimals.  By studying the physical mechanisms
that drive these outflows, and also their interactions with stellar convection,
rotation, pulsation and magnetic fields, we are better able to assess and
quantify the importance of stellar winds to astrophysics in general.  Hence,
understanding stellar winds is one of the fundamental challenges in theoretical
astrophysics.  Clues to the nature of the acceleration of stellar winds should
originate from their underlying base.  In most of the cases (except for a small
number of rapidly rotating giants such as YY Men), that base is represented by
a chromosphere and/or a transition region that is heated to a few 10$^5$\,K,
while cooler giants and supergiants ($(V-R) > 0.8$) manifest winds that emanate
from the extended outer chromosphere.

The net radiative energy flux from the chromosphere---the region above the
photosphere---is more than 10 times greater than that from the entire overlying
transition region and wind.  Chromospheric temperatures are too high to be
explained by radiative heating alone, so there must therefore be some type of
mechanical energy input.  The convective zones constitute the major power
source for such an energy flux, which contributes to the generation of UV,
X-ray and radio emissions from the stellar chromosphere and are thus also
associated with the initiation of the mass outflows known generically as
stellar winds.

In cool evolved stars the chromosphere represents the interface layer between
the photosphere and the wind, and plays a critical role in specifying the
amount of mechanical energy that is dissipated as atmospheric heating and
deposited as the momentum which results in the initiation of the stellar wind.
The chromosphere is key to the mass and energy flux from the entire atmosphere;
it also determines the dynamics and magnetic topology of the overlying layers
which contain the wind, and for its part the wind plays a fundamental role in
the evolution of stars, especially at the latest phases of their
evolution. Therefore, in order to develop realistic theoretical models of
stellar winds from cool evolved stars, we need to understand the nature of the
dominant physical processes, including the mass and energy flow into the
transition region that is positioned between the chromosphere and the wind.
The problem of wind formation can therefore be viewed as a problem of
chromospheric heating, and thence to its acknowledgement as a problem of
fundamental importance for solar and stellar astrophysics.

Chromospheres of cool evolved stars on the red-giant branch (RGB), which are
the principal targets of this review, represent an extreme case of
chromospheres that are observed in cool main-sequence stars. UV observations
suggest two distinct classes of cool giants and supergiants, separated by the
Linsky-Haisch dividing line (Linsky \& Haisch 1979).  On the blueward side of
the dividing line, evolved stars hotter than K2~III (also known as `coronal'
giants) show highly compact chromospheres, transition regions and energetic
coronae~(with $L_{\rm X}/L_{\rm bol}$ up to $10^{-3}$) that extend into
relatively fast ($>$ 100 km~s$^{-1}$), tenuous ($<
10^{-12}~M_{\odot}$~yr$^{-1}$), hot winds ($>1$~MK).  Those stars are
magnetically-active giants; they show variable chromospheric emission and
energetic flares which are observed in the UV, EUV and X-ray regions.  Direct
observations of surface magnetic fields in over 60 active giants have produced
measurements of field strengths that can be as much as 100~G
(Konstantinova-Antova et al.~2012).  The redward side of the dividing line is
characterized by cool, late-K and M-type giants (the `non-coronal' giants);
they show relatively weak transition-region signatures but do have extended
chromospheres ($0.16 - 0.4~R_{\star}$) and massive slow winds (from 10$^{-11}$
to $10^{-6}~M_{\odot}$~yr$^{-1}$), whose typical velocities of $\leq$ 40
km~s$^{-1}$ are considerably smaller than the expected escape velocities at the
stellar surface.  Intermediate between the coronal and the non-coronal giants
are the so-called `hybrid' giants, which show signatures of both compact
chromospheres and pronounced transition regions; those transition regions are
characterized by emission from N~{\sc v}, C~{\sc iv} and Si~{\sc iv}, hot
coronal plasma ($1-20$ MK) observed in X-rays, and hot ($T \sim 1$~MK) tenuous,
fast winds ($>$100\,\kms) and associated mass-loss rates of
$10^{-13}~M_{\odot}$~yr$^{-1}$.

Despite decades of observational and
theoretical studies, however, the physical nature of that atmospheric heating
and associated mass loss is not well understood.
(a) {\it What physical mechanism accounts for the existence of three types of
cool evolved stars?}  (b) {\it What processes control the extent of the
chromospheres in those stars, and their mass-loss rates?} (c) {\it What
physical processes are involved in heating a stellar atmosphere and
accelerating the star's wind?}
To address these questions, any theoretical model of outer atmospheres of cool
stars should describe an intimate relationship between the heating of the
chromosphere, the transition region and the corona, and the heating and
acceleration of the wind.

From that perspective, K5--M giants and supergiants, and in particular those
that are in eclipsing binaries, serve as an ideal laboratory for investigating
how a relatively simple atmosphere that consists of an extended chromosphere of
0.2--1~$R_{\star}$ develops into a slow and massive stellar wind.  In this
review we highlight recent progress in our understanding of chromospheres and
winds from studies of cool evolved stars, and building on previous reviews on
chromospheric heating processes by ({\it inter alia}) Narain \& Ulmschneider
(1990, 1996).  In relation to the atmospheres and winds of cool evolved stars,
we discuss theoretical constraints to heating and acceleration that are derived
from semi-empirical models, and go on to describe the successes that have been
achieved by accounting for atmospheric heating via acoustic waves, but
mentioning also the attendant limitations.  We also highlight recent approaches
to modelling wind acceleration from cool stars from the aspect of
magnetohydrodynamic (MHD) waves, and conclude by discussing possible future
developments in theoretical modelling.

\section{Observational Constraints on the Heating and Acceleration of Stellar
Atmospheres and Winds}

As in the case of the chromosphere of the Sun, the chromosphere of a cool
evolved star consists of highly complex, time-dependent, optically thick,
weakly ionized, magnetized atmospheric layers.  Unlike the solar chromosphere,
however, dynamic chromospheres of non-coronal giants and supergiant stars
transition from quasi-static atmospheres (in the most simplistic case) into
stellar winds.  Models of such complex multi-parameter environments need to
be based on observational constraints.  Various sophisticated multi-dimensional
models of the solar chromosphere that have recently been developed attempt to
describe the chromosphere of the Sun, and---by implication---stellar
chromospheres and winds as well (Suzuki 2007; Hansteen et al.~2010; Airapetian
et al.~2014).  To make those models more realistic, many input parameters are
needed to describe the sources and the specifics of the mechanical energy flux
that is generated within the solar or stellar photosphere.  It also requires a
reference multi-dimensional model atmosphere which is generated by those
models.  There is a variety of methods by which realistic theoretical models
can be constrained.

Extensive observations of samples of G--M giants and supergiants---for
instance, by Carpenter et al.~(1994, 1995), Brown et al.~(1996), Reimers et
al.~(1996), Robinson et al.~(1998), Ayres et al.~(1998, 2003), Ayres (2005),
Dupree et al.~(2005), Harper et al.~(2005, 2013), Harper (2010), P{\'e}rez
Mart{\'{\i}}nez et al.~(2011)--- provide important clues to the thermodynamics
and kinematics of the chromospheres and winds of the various stellar types.
Recent high spectral resolution aperture-synthesis imaging of two supergiants,
Antares and Betelgeuse, has revealed an asymmetry and inhomogeneity of their
chromospheric structures and the existence of a clumpy, cool, outer molecular
shell, the `MOLsphere', extending out to 1.2--1.5 stellar radii (Ohnaka 2013;
Ohnaka et al.~2013).

In considering the atmospheric plasmas of stars, we can divide the observational
constraints into two major categories: energy dissipation requirements, and
momentum deposition requirements.

\subsection{Energy Dissipation Requirements}

In the absence of significant flows, the dissipation of chromospheric energy
due to non-radiative energy source(s) is mostly balanced by radiative cooling.
The observed surface fluxes of the two major contributors, i.e., the Mg~{\sc
ii} and Ca~{\sc ii} emission lines, allow one to define the range of required
heating rates.  Those have been given as $(1-100) \times
10^5$~ergs~cm$^{-2}$~s$^{-1}$ (Linsky \& Ayres 1978; Strassmeier et al.~1994;
P{\'e}rez Mart{\'{\i}}nez et al.~2011).

One-dimensional semi-empirical models of evolved stars represent powerful tools
for constraining the radial profiles of the heating rates that are related to
the deposition of energy throughout the atmosphere.  This class of model was
inspired by time-independent 1-D semi-empirical models of the solar
chromosphere developed by Vernazza et al.~(1976, 1981) and Fontenla et
al.~(1990, 2002); they were designed to reproduce the temporally and spatially
averaged UV line profiles and fluxes.  Semi-empirical models provide a
quantitative characterization of the radial profiles of temperature, electron
density, neutral hydrogen density and turbulent velocity across the atmospheres
of evolved stars.  This type of model was developed for a number of evolved
stars, such as giants like $\alpha$~Boo, $\alpha$~Tau, and $\beta$ Cet, and for
various supergiants, including the eclipsing supergiant 31~Cyg (Ayres \& Linsky
1975; Eriksson et al.~1983; McMurry~1999; Eaton 2008).  A chromospheric
model for $\alpha$ Tau developed by McMurry (1999) suggests that the
temperature rises throughout the chromosphere up to 100,000\,K at about
0.2~$R_{\star}$.  At the same time, the chromosphere transitions into a wind
within one stellar radius, suggesting that the atmosphere therefore undergoes
acceleration between 0.2 and 1~$R_{\star}$.  {\it FUSE} observations of various
non-coronal giants show the presence of C~{\sc iii} and O~{\sc vi} lines,
indicating hot plasma with temperatures up to 300,000\,K.  Plasma at such high
temperatures occupies low volumes and appears to be mostly at rest with respect
to the photosphere in stars that have winds of low escape velocities,
indicating that the plasma should be magnetically confined (Ayres et al.~2003;
Harper et al.~2005; Carpenter \& Airapetian 2009).

Recent detections of surface magnetic fields for some G--M giants and
supergiants suggest that surface magnetic fields could be an important
contributor to the thermodynamics of the outer chromosphere (Auri\'ere et
al.~2010; Konstantinova-Antova et al.~2010, 2012).  The observed field
strengths vary from 0.5 to 1.5~G in late-type giants and increase to 100~G in
early-type coronal giants.  Rosner et al.~(1995) suggested that as stars evolve
toward the giant phase, their magnetic topology transitions from closed
magnetic configurations to predominantly open ones; the latter allow massive,
non-coronal winds to be supported.  If the magnetic field is non-uniformly
distributed over the stellar surface, the associated radial profiles in the
atmosphere can be determined by assuming that the magnetic pressure inside an
untwisted (purely longitudinal) flux tube, ${B^2}/{8\pi}$, is balanced by the
gas pressure of the surrounding non-magnetic atmosphere, $P_{\rm ext}$. This
suggests that the plasma pressure inside the tube is smaller than the magnetic
pressure of the plasma, $\beta = {7 n_9 T_4}/{B_1^2}$, where
$n_9=n/10^9$~cm$^{-3}$, $T_4=T/10,000$\,K, and $B_1=B/10$~G.  For typical
chromospheric conditions of $n_9 \sim$ 1 and $T_4\sim$1,
the plasma-$\beta$ becomes less than 1 at B $\geq$ 50 G.  Observations in the
vicinity of active regions on the Sun that are represented by plages indicate
magnetic fields of a few hundred Gauss at chromospheric densities and
temperatures; the force balance between the magnetic and plasma pressures can
therefore be described satisfactorily by the thin flux-tube approximation
(Rabin 1992; Gary 2001; Steiner 2007; Judge et al. 2011).  The vertical profile of the
chromospheric magnetic field can therefore be determined as \be B_z(z) =
\sqrt{8~\pi~P_{\rm gas}} . \label{ch5:eq:1} \ee Once the magnetic field is known, the profile
of the Alfv\'en velocity, $V_A$, can be calculated throughout the chromosphere
as \be V_A= \frac{B_z(z)}{\sqrt{4~\pi~\rho(z)}} , \label{ch5:eq:2} \ee where $\rho(z)$ is the mass density.

Since the photospheres of giants and supergiants are convective and dense,
photospheric footpoints of longitudinal magnetic fields are forced to follow
the convective motions within the photosphere. The motions of magnetic field
lines with a frequency of the inverse turnover time of a stellar granule,
$\nu_A = H_p/V_c$, with $H_p$ as photospheric pressure scale height and $V_c$
as convective velocity, are able to excite MHD waves along or across the
magnetic flux tube, including torsional or transverse Alfv\'en waves (Ruderman
et al.~1997).  Torsional Alfv\'en waves (Alfv\'en 1942) represent linearly
incompressible azimuthal perturbations of the plasma velocity (linked to the
azimuthal perturbations of the magnetic field) that, unlike compressible waves
(such as longitudinal MHD waves), do not disturb the plasma density.  Although
lfv\'en waves were predicted in 1942, it is only relatively recently that
researchers have reported the observational detection of them in the solar
chromosphere and corona (Tomczyk et al.~2007; De Pontieu et al.~2007; Jess et
al.~2009).

Alfv\'en waves launched from the stellar photosphere propagate upward into a
gravitationally stratified atmosphere and are subject to reflection from
regions of high gradients of Alfv\'en velocity if the wave frequency,
$\nu_{A}$, is less than the critical frequency, $\nu_{\rm crit} = dV_A/dz$
(Heinemann \& Olbert 1980; An et al.~1990).  The interaction of downward
reflected Alfv\'en waves with upward propagated ones can ignited a turbulent
cascade of Alfv\'en waves in the lower solar atmosphere and provide a viable
source for the solar coronal heating and stellar wind heating in the open field
regions (Matthaeus et al. 1999; Cranmer \& Ballegooigen 2005; Cranmer 2011).

 Reflection of Alfv\'en waves can play an important role in driving slow and
 massive winds from giants and supergiants (An et al.~1990; Airapetian et
 al.~1998, 2000, 2010; Suzuki 2007; Cranmer 2011).  The radial profile of the
 critical frequency therefore provides important information about the role of
 the heating and momentum deposition of Alfv\'en waves in the atmosphere.  The
 critical Alfv\'en frequency can be calculated directly from a semi-empirical
 model by differentiating the Alfv\'en velocity profile given by
 Eq.~(2).  Airapetian et al.~(2013) have applied this procedure
 to the case of a K5 giant ($\alpha$ Tau) using the semi-empirical model of
 McMurry (1999).  They showed that the Alfv\'en velocity gradient reaches its
 maximum at 0.21 $R_{\star}$.  Figure~1 suggests that waves at
 frequencies less than 0.6 mHz are trapped in the chromosphere of $\alpha$~Tau
 (left panel).  They also applied the same technique, based on the
 semi-empirical model developed by Eaton (2008), to calculate the gradient of
 the Alfv\'en velocity for the K5 supergiant primary in 31~Cyg.  The right
 panel of Fig.~1 shows the vertical profile for $dV_A/dr$ in the
 chromosphere of 31~Cyg, and suggests that waves at $\nu \leq$ 3 nHz should be
 trapped in the first 10 stellar radii.

The magnetic field and the Alfv\'en velocity profile can also be probed by
using the Poynting theorem (Jackson 1999): \be \frac{\partial{W}}{\partial{t}}
+ \vec{\nabla} \vec{S} = - \vec{E} \vec{j} , \ee where $W =
\frac{1}{8\pi}(E^2+B^2)$ is the electromagnetic energy density and
$\vec{S}=\frac{1}{4\pi}~\vec{E}\times \vec{B}$ is the Poynting vector of the
energy source.  $\vec{S}$ represents the Poynting flux of Alfv\'en waves
launched from the photosphere.  For a steady-state chromosphere,
$\frac{\partial{W}}{\partial{t}}$ = 0 and $\vec{S}$ has only an upward
component $S_z$.  We thus obtain \be \frac{dS_z}{dz} = - <q> , \label{ch5:eq:4}
\ee where $<q>$ is the time-averaged heating rate at a given height, $z$ (see
also Song \& Vasyliunas 2011).

The heating rate of the plasma can be derived from the energy equation for a
steady-state chromosphere where the heating rate is balanced by the thermal
conductive and radiative cooling rates, referred to as $L_{\rm cond}$ and
$L_{\rm rad}$, respectively.  It is found that \be <q> = L_{\rm cond} + L_{\rm
  rad} . \label{ch5:eq:5} \ee

In a stellar chromosphere, $T<0.5$~MK; the thermal conduction
time is therefore much longer than the radiative cooling time, and the thermal
conduction cooling term can be safely neglected.  Consequently, the radial
profile of the observationally derived cooling rates provides direct clues
about the profile of the Poynting flux of the heating energy source.  Detailed
information about the radial profiles of the chromospheric magnetic field and
the Alfv\'en velocity can be obtained if it is assumed that Alfv\'en waves are
the major source of the chromospheric heating.  The observations of
non-thermally broadened chromospheric lines also imply that Alfv\'en waves may
be the dominant source of energy and wind acceleration in cool giants and
supergiants; further relevant discussions have been given by Airapetian et
al.~(2010) and by Cranmer \& Saar (2011).  This type of incompressible
transverse wave can be directly excited, presumably through the shuffling or
twisting of magnetic flux tubes by well developed magneto-convection in stellar
photospheres (Ruderman et al.~1997; Musielak \& Ulmschneider 2002).  Recently,
Morton et al.~(2013) presented observational evidence that, in the solar
photosphere, incompressible waves can be excited by the vortex motions of a
concentration of magnetic flux.

The energy flux of Alfv\'en waves excited at the photosphere is defined by the
$z$-component of the Poynting vector $\vec{S} = \frac{1}{4\pi}~
\vec{E}~\times~\vec{B}$.  By applying Ohm's law, $\vec{E} = \eta \vec{j} -
\vec{V}\times\vec{B}$, Ampere's law, $\vec{j} = \frac {1}{4\pi} \vec{\nabla}
\times \vec{B}$, and using vector identities, we can write the upward Poynting
flux in Alfv\'en waves as

\be
\vec{S} = \frac{1}{4~\pi} [\vec{V}~B^2 - \vec{B}(\vec{V} \cdot\ \vec{B})] +
\frac{\eta}{4\pi}(\vec{\nabla} \times \vec{B}) \times {\vec{B} .} \label{ch5:eq:6}
\ee

If we further assume the existence of the azimuthal component only of the
velocity of footpoint motions, $V_{\phi} \neq 0$, i.e., that there are no
vertical motions in the photosphere (so $V_z = 0$), and if we represent the
total magnetic field as the sum of the background longitudinal flux-tube
magnetic field $B_z$ plus the perturbed field $\delta{B}$ due to Alfv\'en
waves, we obtain the $z$-component of the upward Poynting flux as

\be
S_z = -\frac{1}{4~\pi}~B_z~V_{\phi}~{\delta B} - \frac{\eta}{4~\pi}
{\delta B} \frac{\partial{\delta B}}{\partial{z}} . \label{ch5:eq:7}
\ee

For high magnetic Reynolds numbers, ${\rm Re}_m = \frac{V_A L}{\eta}$ ($\eta$
is the magnetic diffusivity), in the stellar chromosphere ($>$ 10, the
second term in Eq.~(7) can be neglected with
respect to the first term.  Then, following the Wall\'en relation
$\delta V_A = \delta \frac{B}{\sqrt{4\pi\rho}}$ and assuming that waves are
incompressible (so $\delta \rho$ = 0), we obtain

\be
\frac {\delta {V}}{V_A} = \frac{\delta{B}}{B_z} . \label{ch5:eq:8}
\ee

This assumption is valid until Alfv\'en waves become strongly non-linear and
convert a significant fraction of their energy into longitudinal waves (Ofman
\& Davila 1997; Suzuki 2007; Airapetian et al.~2014).  Substituting $\delta B$
from Eq.~(8) into Eq.~(7), we obtain the Poynting flux as

\be
S_z = \rho~<\delta V^2> V_A . \label{ch5:eq:9}
\ee
Furthermore, when combining Eqs.~(4), (5), and (9), we obtain the following:

\be
\frac{d}{dz}(\rho~<\delta V^2> V_A) = - L_{\rm rad}(z), \label{ch5:eq:10}
\ee

Equation~10 relates the thermodynamic quantities such as the plasma density,
turbulent velocity and the radiative cooling rates, which are obtained from
semi-empirical models, to the hitherto unknown vertical profile of the Alfv\'en
velocity.  Equation~10 can be rewritten as:

\begin{equation}
e_A~\frac{dV_A}{dz} + \frac{de_A}{dz}~V_A = - L_{\rm rad}, \label{ch5:eq:11}
\end{equation}
where  $e_A=\rho~<\delta V^2>$ is the energy density of Alfv\'en wave energy.

Once $V_A$ is known, the profile of the magnetic field throughout the
chromosphere can be determined. {\it Hence, the knowledge of $V_A$ and
  subsequent retrieval of $B_z(z)$ represents the missing link between
  thermodynamic-based semi-empirical models and MHD-based theoretical models of
  chromospheres and winds.} This last equation allows us to determine the range
of critical frequencies at which Alfv\'en waves become reflected from regions
where the Alfv\'en velocity gradient is at a maximum.

Comparing the magnetic-field profiles derived from Eq.~(11) with the one
obtained from Eq.~(1) enables us to determine the degree of deviation of
the magnetic field in a chromosphere from the longitudinal (untwisted) magnetic
field, thus allowing us to constrain the value of the azimuthal magnetic field.
The magnetic-field profile in the chromosphere of $\alpha$ Tau as derived
decreases with height at the rate of a super-radial expansion factor, $f(r)$.
Then, the magnetic field varies with the height, $r$, as $B(r)\sim
f(r)/r^2$, which is less steep than the profile obtained by Kopp \& Holzer
(1976) for solar coronal holes.

The next generation of semi-empirical models of evolved stars should therefore
combine high-resolution spectroscopic and spatial information.  Eclipsing
binaries offer a unique opportunity to derive geometric constraints on the
observed chromospheres and their winds (Eaton et al.~2008). Another promising
approach utilizes high spatial-resolution interferometric observations of
various giant and supergiant stars.

\subsection{Momentum Deposition Requirements}

The momentum deposition from non-radiative energy sources into stellar winds
should explain the observed energy fluxes.  An estimate of the energy flux that should be added
in order to drive a steady-state wind above the atmospheric base
can be derived from the energy equation for a steady-state wind in a
single-fluid hydrodynamic approximation:

\be F_{\rm wind}=\frac{(-\dot{M})}{4\pi~R_{\star}^2}~(V_{\infty}^2 +
V_{\rm esc}^2). \label{ch5:eq:12}
\ee
$\rho$ is the wind mass density, $V_{\infty}$ is the wind terminal velocity,
which is reached at radius $R_{\star}$, and $V_{\rm esc}$ is the escape velocity of
the star at its surface (Holzer 1987).  Detailed examination of chromospheric emission lines
of Fe~{\sc ii}, O~{\sc i} and Mg~{\sc ii} indicate that the wind from a giant
appears to originate near the base of the chromosphere and continues to
accelerate throughout the entire chromospheric region (Carpenter et al.~1995).
It is therefore assumed that the wind reaches its terminal velocity within one
stellar radius, i.e., $R_0=R_{\star}$.

Table~1 presents radii, $R_{\star}/R_{\odot}$, escape velocities, $V_{\rm
esc}$, wind velocities, $V_{\infty}$, and mass loss rates, $\dot{M}$, for
selected stars (Robinson et al.~1998; Eaton 2008; Harper~2010; Neilson
et al.~2011).  One can see that a significant portion (from 65 to 95\%) of the
mechanical energy is required to lift the plasma beyond the gravitational well
of the star and form the wind.

The energy flux required to generate the winds from $\alpha$~Tau, $\alpha$~Ori
and 31~Cyg are $2.8\times 10^3$, $2.2 \times 10^5$ and $2.5 \times
10^5$~ergs~cm$^{-2}$~s$^{-1}$.  That suggests that in non-coronal giants,
$F_{\rm wind}$ is only a few percent of the outer atmospheric heating rate
and therefore over 90\% of the energy flux of the mechanical source is
trapped in the chromosphere while only a small fraction leaks out to accelerate
and heat the wind. For supergiants the wind is initiated in much higher
regions of the extended atmosphere; it reaches its terminal velocity before
leaving the chromosphere (see Carpenter \& Robinson 1997).

Another requirement for models of stellar winds arises from the constraints to
the terminal wind velocities and mass-loss rates.  For example, non-coronal
giants and supergiants, including those in \za~systems, show evidence of the
presence of slow (a few tens of km~s$^{-1}$) and massive winds. Models of
steady-state, spherically-symmetric winds suggest that in order to produce the
low terminal velocities (ones that are less than half the surface escape
velocities) and the high mass-loss rates, most of the energy and momentum
should be deposited below the sonic point, while momentum addition beyond the
sonic point should produce the fast and tenuous solar-like winds (Hartmann \&
MacGregor 1980; Holzer 1987).

\subsection{Constraints from Atmospheric Turbulence and Flows}

High-resolution spectroscopy of many cool giants and supergiants has
demonstrated that the profiles of optically thin UV lines of C~{\sc ii}],
Si~{\sc ii}, Si~{\sc iv} and C~{\sc iv} that arise in the plasma that
constitutes the chromosphere and transition regions in those stars show direct
evidence of strong turbulence.  The C~{\sc ii}] line is formed at temperatures
between 5,000 and 10,000\,K, and is optically thin in all of the stars
observed.  It has an intrinsically narrow profile, so the line width primarily
reflects the Doppler broadening; the enhanced wings of the line profiles can be
simulated by a single Gaussian profile (Carpenter et al.~1991).  The deduced
turbulent velocities range from 24\,\kms~for the K5 giant $\alpha$~Tau to
35\,\kms~for the M1.5 supergiant $\alpha$~Ori, thus suggesting supersonic
turbulence in their chromospheres; see also discussions by Cuntz (1997),
Robinson et al.~(1998) and Cuntz et al.~(2001).  Many of the UV line profiles
can be fitted by narrow and broad Gaussian components, as discussed by Wood et
al.~(1997), Dupree et al.~(2005) and Eaton (2008). For example, fitting the
profiles of C~{\sc ii}] in the M3.5\,III star $\gamma$ Cru requires a
two-component Gaussian model with FWHM = 27 and 42\,\kms, and for $\alpha$~Ori,
19 and 48\,\kms~(Eaton 2008).

Hybrid and coronal giants show much larger non-thermal velocities in transition
region lines reaching 200 km~s$^{-1}$, and also show narrow and broad Gaussian
components.  Moreover, non-thermal broadenings of UV lines observed in
quiescent spectra of coronal giants as well as active dwarfs including the Sun
tend to increase with temperature (Ayres et al. 1998; Pagano et al. 2004; Peter
2006).

Our Sun also shows non-thermally broadened, two-component Gaussian-shaped,
red-shifted UV lines that form in the transition region (Peter 2006).  That
study does not confirm the idea suggested by Wood et al.~(1997) and Pagano et
al.  (2004) that the broad component of UV lines is heated by microflares. In
contrast, it used the spectrum of the Sun's integrated disk to demonstrate that
the broad components indicate the structure of its chromosphere. Carpenter \&
Robinson (1997) suggested that broad components may be a misleading description
of the physics of a stellar chromosphere; instead of fitting the profiles by
two Gaussian curves, they explain the enhanced wings by a signature of
large-scale turbulence, which is anisotropically distributed along the line of
sight and is directed preferentially either along, or perpendicular to, the
radial direction.  Airapetian et al.~(1998, 2000, 2010) developed a 2.5-D MHD
model of a stellar wind in which supersonic turbulent motions that are
responsible for non-thermal broadening in the UV lines can be attributed to
unresolved motions of the upward propagating non-linear Alfv\'en waves.  The
anisotropy formed by a large-scale longitudinal (open) magnetic field in an
atmosphere can contribute to the formation of broad wings such as those
observed in evolved giants and supergiants. Another important feature of the
outer atmospheres of cool stars like the primary of 31 Cyg is the clumping of
the gas, as proposed by Eaton (2008) and discussed further by Harper (2010).

Optically thin chromospheric lines from these stars and the Sun also exhibit a
net redshift. This red-shift is indicative of downward motions of a
few~\kms\ in non-coronal giants, and up to 30\,\kms\ in coronal giants as
observed in lines from the chromosphere and transition region (Peter \& Judge
1999; Ayres et al.~1998; Robinson et al.~1998; Doyle et al. 2002; Eaton
2008). It is interesting to point that Doyle et al. (2002) has observationally
shown that the red Doppler shifts in the solar emission lines of O~{\sc III},
O~{\sc IV} and Ne~{\sc VIII} show short (a few minutes) variability. This time
scale is comparable to the characteristic period of MHD waves observed in the
solar chromosphere with the period of a few minutes (see Morton et
al. 2012). Assuming that MHD waves are excited by magneto-convective motions
with the turbulent turnover time of a convective cell, $\tau = L_g/V_{conv}$,
we can expect a variation of red-shift shifts in chromospheric lines of red
giants at the time scale of a few days.  Judge \& Carpenter (1998) concluded
that such turbulent motions indicate downward propagating non-linear waves in
the chromosphere.  3-D MHD models of the transition region in the Sun also
interpret the observed emission-line red-shifts in terms of downward
propagating compressive waves (Hansteen 1993), or material heated within
low-lying transition-region loops that later cool and fall down into the
chromosphere (Guerreiro et al.~2013).  This process can be very important in
the chromospheric dynamics and the energy balance, and therefore, provide a
clue to the solar and stellar atmospheric heating. Can those models be
incorporated into describing the mechanisms for heating chromospheres and
winds?  Thus, non-thermal broadening and redshifts of chromospheric lines
represent one of the major signatures of chromospheric heating that need to be
addressed by any viable theoretical model.

\section{Acoustic Heating: Successes and Limitations}

\subsection{Two-Component Chromosphere Models}

Following previous reviews by, e.g., Narain \& Ulmschneider (1990, 1996),
including references therein, particularly the work by Schrijver (1987)
and Rutten et al.~(1991), it has become obvious that from a general point
of view stellar chromospheres can be considered as consisting of
acoustically heated and magnetically heated components.  The general
heating rate, proliferated to stellar chromospheres including those
of non-coronal giants and supergiants, can thus be expressed as
\be
F \ = \ F_{\rm ac}(T_{\rm eff},g_{\star},Z) + F_{\rm mag}(T_{\rm eff},g_{\star},Z,P_{\rm rot}) ,
\ee
where $F_{\rm ac}$ is the acoustic heating rate and $F_{\rm mag}$ is the
magnetic heating rate.  Both increase as function of stellar effective
temperature $T_{\rm eff}$, increase with decreasing stellar surface gravity
$g_{\star}$ (i.e., by a few orders of magnitude between main-sequence stars and
low-gravity supergiants), and also show some dependence on the stellar metallicity
$Z$, which is however much less important than the impact of $g_{\star}$.

The magnetic heating rate also depends---at least in a statistical sense---on
the rotation rate of the star, i.e., it is lower for slow rotating (i.e.,
older) stars, especially giants and supergiants, which have evolved away from
the main-sequence (i.e., Schrijver 1993).  This behaviour has profound
consequences for the resulting amounts of chromospheric emission (e.g.,
Skumanich 1972; Noyes et al. 1984; Simon et al. 1985; Strassmeier et al. 1994),
as identified in multiple spectral regimes, including detailed observations of
Ca~{\sc ii} and Mg~{\sc ii}.  The latter have been interpreted based on
empirical, semi-empirical and theoretical concepts, including statistical
relationships, and have also been utilized to decipher information about the
thermodynamic, (magneto-)hydrodynamic and radiative properties of stars at
different ages and evolutionary status.  Detailed analyses focused on
non-coronal giants have been described by Schrijver (1993), Schrijver \& Pols
(1993), and others.  Furthermore, theoretical studies for the linkage between
stellar atmospheric and wind properties, on the one hand, and the evolution of
stellar dynamos, on the other hand, which actually constitute the physical
reason for the fundamental changes, as the transition from coronal
main-sequence stars to non-coronal giants, were given by MacGregor \&
Charbonneau (1994) and Charbonneau \& MacGregor (1995).

Another pivotal aspect concerns the study of one-component (acoustic only) and
two-component (i.e., acoustic and magnetic) chromospheric heating simulations
itself.  Detailed theoretical models for stars of different spectral types as
well as non-coronal giants have been given by Ulmschneider (1989), Buchholz et
al.~(1998), Cuntz et al.~(1998, 1999), and Fawzy et al.~(2002).  For example,
Buchholz et al.~(1998) presented 1-D time-dependent models of acoustically
heated chromospheres for main-sequence stars between spectral type F0~V and
M0~V and for two giants akin to spectral type K0~III and K5~III.  The emergent
radiation in Mg~{\sc ii} {\it h}+{\it k} and Ca~{\sc ii} H+K was calculated and
compared with observations.  They found good agreement, over nearly two orders
of magnitude, between the time-averaged emission in these lines and the
observed basal flux emission, which had been suspected to be due to nonmagnetic
(i.e., acoustic) heating operating in all late-type stars.  The authors pointed
out that the results obtained clearly support the idea that the `basal
heating' of the chromospheres of late-type stars, including non-coronal giants
(see Fig.~2) is due to acoustic waves.  The latter result has been disputed
in the meantime (Judge \& Carpenter 1998) from the view of that acoustic wave
heating, in the context of existing models, is unable to explain the amounts of
observed turbulence (see Sect.~2.3).

\subsection{Possible Relevance of Acoustic Waves for Winds from Cool Evolved
Stars}

An important feature of acoustic waves in stellar atmospheric environments,
including the chromospheres of evolved giants and supergiant stars, is that
they dissipate the lion's share of their mechanical energy flux fairly close to
the stellar photospheres; see, e.g., the study of Ulmschneider (1989) that
compared the dissipative behaviour of acoustic waves in giants and dwarfs of
identical effective temperatures.  This key result about the dissipative
properties of acoustic waves in cool evolved stars, implying that acoustic
waves cannot support the massive winds of those stars, has also been discussed
in the broader context of proposed stellar mass loss mechanisms by Holzer
(1987).  Thus, this entails an additional justification for the exploration of
Alfv\'en wave driven winds for non-coronal evolved stars as opposed to
nonmagnetic mechanisms.  Alfv\'en wave driven wind models were pursued by
Hartmann \& MacGregor (1980), Hartmann \& Avrett (1983), and more recently by,
e.g., Airapetian et al.~(2014).  A detailed study by Cuntz (1990), focused on
$\alpha$~Boo (K2~III) and based on adequate acoustic wave frequency spectra,
yielded acoustically initiated mass loss rates on the order of $10^{-13}$ to
$10^{-15}$ $M_\odot$~yr$^{-1}$, which are more than a factor of 10$^4$ below
the observationally established limit.

The decisive shortcoming of acoustic waves consists in that they fail to
dissipate their mechanical energy over a significant distance that is
comparable to the height of the critical point (Holzer \& MacGregor 1985; Cuntz
1990; Judge \& Stencel 1991).  This property thus leads to absurdly low mass
loss rates due to acoustic waves.  Previous relatively positive assessments
about the ability of acoustic waves to drive mass loss from evolved stars were
given by Pijpers \& Hearn (1989).  They explored a simple stellar wind model
loosely guided by the stellar parameters of $\alpha$~Ori, which led to mass
loss rates in the range between 10$^{-8}$ and 10$^{-4}$ $M_\odot$~yr$^{-1}$,
depending on the wave parameters, which at first sight appear to be promising.
However, this type of model must be ruled inapplicable as it is based on the
assumption of time-independent and linear behavior of the acoustic waves,
excluding, for example, the possibility of shock formation. Thus the results
that were obtained then carry little meaning.

Nevertheless, acoustic waves are still expected to be relevant for initiating
mass loss from non-coronal giants and supergiant stars as they are able to
increase the thermal pressure and density scale heights in the spatially
extended chromospheres owing to their dissipative behaviors (i.e., heating
and transfer of wave pressure), see Buchholz et al.~(1998) and references
therein.  Another possible contribution of nonmagnetic processes is the
initiation of turbulent pressure that may help to lift matter closer
to the critical point of the stellar wind as pointed out by, e.g.,
Suzuki (2007).  Detailed assessments, in the framework of 1-D models, of
acoustically generated turbulence in chromospheric models of $\alpha$~Tau
show however that the synthetic results are significantly lower than
those obtained by GHRS observations (Judge \& Cuntz 1993).  However, it is
still an unsettled debate (see Judge \& Carpenter 1998; Cuntz et al. 2001)
whether this discrepancy is due to the unrealistic 1-D assumption of
existing theoretical models, or whether it is more profoundly enshrined
in the basic physics of acoustic processes.  Preliminary 3-D models of
convectively initiated turbulence in case of $\alpha$~Ori have been
given by Freytag et al.~(2002).

\section{MHD Wave Driven Heating and Wind Acceleration}

\subsection{Energy Dissipation Due to Alfv\'en Waves: A Source of Chromospheric
 Heating}

Magnetic heating mechanisms for solar and stellar chromospheres have been
targeted in numerous reviews, including those by Narain \& Ulmschneider (1990,
1996).  Two major types of heating mechanisms have generally been proposed,
which are broadly classified as AC (i.e., alternating current, such as MHD wave
dissipation) and DC (i.e., direct current, such as magnetic-field dissipation
through magnetic reconnection).  In this review we focus on the progress
regarding AC heating processes and their observational signatures.  MHD-wave
heating can be driven by two types of waves: compressible longitudinal MHD
waves (slow and fast magneto-sonic waves), and incompressible transverse waves
(i.e., torsional Alfv\'en waves).  According to Ulmschneider et al.~(2001) and
Musielak \& Ulmschneider (2002), the energy fluxes of longitudinal and
transverse waves in cool evolved stars are comparable, and are of the order of
$10^7 - 10^8$~ergs~cm$^{-2}$~s$^{-1}$.  That amount of energy generated by
waves in stellar photospheres of cool giants is sufficient to account for the
observed cooling rates together with the energy needed to drive winds (see
Sect.~2.1.

Torsional Alfv\'en waves have been suggested as a likely source for the heating
of the solar chromosphere and corona (Osterbrock 1961; Hollweg 1973, 1978;
Heinemann \& Olbert 1980; Holzer et al.~1983; Cranmer \& Ballegooigen~2005;
Mathioudakis et al.~2013).  That approach has also been extended to the outer
atmospheres of cool evolved stars (Hartmann \& MacGregor 1980; Holzer et
al.~1983; Hartman \& Avrett 1984; Suzuki 2007, 2013; Cranmer 2008, 2009).

Alfv\'en waves can damp energy in solar and stellar atmosphere through a number
of mechanisms.  For example, in closed magnetic structures resonant absorption
mechanism may become efficient, while in closed and open structures energy
dissipation through the cascade due to Alfv\'en wave turbulence or mode
conversion may become efficient (Davila 1987; Matthaeus et al.~1999; Cranmer \&
Ballegooigen 2005; Suzuki \& Inutsuka 2005; Mathioudakis et al.~2013;
Airapetian et al.~2014).  Studies by An et al.~(1990), Barkhudarov (1991),
Rosner et al.~(1991), Velli (1993), MacGregor \& Charbonneau (1994) and
Charbonneau \& MacGregor (1995) concluded that as waves propagate in a
gravitationally-stratified atmosphere they may become subject to reflection
from atmospheric regions where the gradient in the Alfv\'en velocity is
comparable to, or greater than, the Alfv\'en wave frequency.

While those studies clearly point to the possible importance of magnetic-wave
pressure in chromospheric heating, they suffer from the restrictive nature of
the linearity assumptions as well as from the fact that they are not
consistently solving the relevant MHD equation involving the magnetic field,
density and velocity.  Although the linear treatment of winds in cool, luminous
stars has shown that MHD turbulence can be important for driving the winds,
those models are incapable of examining properly the wave dissipation---a
critical part of the mechanism.  Furthermore, because the entire set of
non-linear time-dependent MHD equations is not solved consistently in those
studies, there is the possibility that important physical effects are neglected
or overlooked.  An important example is that the coupling between the azimuthal
and radial components of the velocity and magnetic fields are only treated in a
linear approximation.  Recent models by Suzuki (2007, 2013) treat
self-consistently the dissipation of Alfv\'en waves in forming stellar
chromospheres and coronal layers that expand into winds, but they assume that
Alfv\'en waves are launched from a fully-ionized photosphere---an approximation
that is not applicable, since the degree of ionization in giant and supergiant
photospheres is less than 10$^{-5}$, and is less than 1 in most
parts of the chromosphere. That approach therefore overlooks a wide range of
physical effects of wave dissipation in a partially-ionized magnetized plasma.

In a weakly ionized medium such as a stellar chromosphere, where there is
varying collisional coupling between ions and electrons throughout, new effects
can appear, such as the non-ideal Hall effect or ambipolar diffusion.  In
astrophysics, ambipolar diffusion usually refers to the decoupling of neutral
and charged particles in a plasma.  If both electrons and ions are magnetized
(or frozen-in into the magnetic field), neutral particles do not `feel' the
magnetic field and will slip through it.  Neutral particles then drag ions with
them via ion--neutral collisions, introducing an electric field perpendicular
to the magnetic field lines. The Hall effect occurs when electrons are
magnetized but ions are not. In such a case the Hall electric field results
from the drift velocity of electrons with respect to ions, because the two
kinds of charged particles respond differently to collisions from the neutral
particles.  In solar and stellar atmospheres the ambipolar diffusion
perpendicular to the magnetic field is many orders of magnitude greater than
the classical Spitzer diffusion along the magnetic field.  The collisional
coupling between charged particles and neutral gas is therefore a fundamental
process in weakly ionized and strongly magnetized solar and stellar
chromospheres and winds (Piddington 1956; Osterbrock 1961; Hartman \& MacGregor
1980; Holzer et al. 1983).  Many recent studies (Goodman 2000, 2004; De Pontieu
et al.~2001; Khodachenko et al.~2004; Leake et al.~2005; Krasnoselskikh et
al.~2010; Soler et al.~2013; Tu \& Song 2013) have shown that in the solar
chromosphere, which is a weakly ionized and magnetized atmosphere, the effect
of ion--neutral collisions becomes significant in dissipating the electric
currents introduced by MHD waves.

Airapetian et al.~(2014) recently showed that the effects of ambipolar
diffusion may also play a crucial role in the chromospheres of cool giants.  It
is known that the photospheres of cool giants and supergiants are characterized
by well-developed magneto-convection with characteristic velocities up to
10\,\kms~(Gray 2008; Chiavassa et al.~2010).  Interaction of such motions with
open magnetic fields may excite longitudinal and transverse MHD waves.  What
happens is that transverse Alfv\'en waves cause periodic fluctuations of plasma
motions perpendicular to the magnetic field lines and thereby induce an
electric field, $\vec{E}= \vec{\delta V} \times \vec{B}/c$, in the reference
frame of the plasma.  This induced electric field generates electric currents.
The dissipation of the currents induced by the Alfv\'en waves perpendicular to
the magnetic field provides an efficient source for converting the kinetic
energy of convection into electrical energy.  Working from the approach of
Braginskii (1965), De Pontieu \& Haerendel (1998), De Pontieu et al.~(2001),
Goodman (2004), Leake et al.~(2005), Leake \& Arber (2006) and Tu \& Song
(2013), Airapetian et al.~(2014) developed MHD models of Alfv\'en-wave heating
for a partially ionized plasma in a solar or stellar chromosphere.  A key
component of those types of models is the inclusion and self-consistent
calculation of the anisotropic electrical conductivity tensor.

To describe the role of ambipolar diffusion and the Hall effect in the
chromosphere of a cool evolved star we need to know how the thermodynamic
parameters of the atmosphere vary with height. Those parameters are provided by
semi-empirical models of the chromosphere, complemented by the magnetic-field
profiles (see Sect.~2.1).  In the following, as an example we
focus on McMurry's semi-empirical model for $\alpha$~Tau.  We also consider
that the chromosphere has a longitudinal but vertically diverging magnetic
field, $B_z = B_0~(R_{\star}/r)^2$, where $B_0$ = 20 G.  The upper left panel
of Fig.~3 shows the radial profile of the chromospheric
temperature, electron number density and N$_H$ (the sum of the neutral and ion
densities).  The upper right panel presents the radial profile of the neutral
fraction, $n_H/n_{\rm tot}$, and the lower one shows the vertical profile of
the plasma beta, $\beta = P_{\rm gas}/P_{\rm mag}$.  The magnetic pressure
appears to be dominant in most of the chromosphere.

In a partially ionized plasma consisting of electrons, protons and neutral
particles, the motions of charged particles are strongly affected by
electron-neutral and ion--neutral collisions. In such a highly collisional
plasma of a stellar chromosphere, neutral hydrogen, electrons and protons are
efficiently coupled, thus constituting a single fluid if the electron-ion
collision frequency is greater than the characteristic frequency of the waves
propagating through the medium. This approximation is valid for low-frequency
Alfv\'en waves (less than 1 Hz) in a chromosphere.  Electrons and
protons gyrate along the magnetic field with characteristic frequencies
of $f_{ce} = 2.8 \times 10^6$~$B$ for electrons and $f_{cp} = 1.52 \times
10^3$~$B$ for protons.  For a longitudinal magnetic field of 100~G, such
frequencies are in the range $10^5-10^8$ Hz.

Unlike fully ionized plasmas, the photosphere and chromosphere of a red giant
contain a large fraction of neutral species that collide with electrons and
ions at frequencies $f_{en}$ and $f_{in}$, respectively.  They are given as \be
f_{en}=1.95 \times 10^{-10}~n_H~\sqrt{T} , \ee \be f_{pn}=7.87 \times
10^{-11}~n_H~\sqrt{T} , \ee and are found to vary within a frequency range of
$10^2-10^4$~Hz.  Consequently, over the entire range of relevant heights in the
atmosphere, the electron and ion gyrofrequencies exceed the proton-neutral and
electron-neutral collision frequencies.  This means that both electrons and
ions are magnetized throughout the chromosphere.  It also suggests that the
Hall effect is negligible for chromospheric conditions, but---according to
Goodman (2000)---it is important in the lower parts of the solar chromosphere.
Mitchner \& Kruger (1973) showed that in plasmas where the magnetization of
electrons, $M_e$, and ions, $M_i$ (i.e., the ratio of electron or ion
gyrofrequency to the total collision frequency of electrons or ions with
neutral particles), becomes greater than 1, the plasma conductivity becomes
anisotropic.  First, this requires that the Spitzer resistivity, which is
parallel to the magnetic field, needs to be modified from its fully ionized
value from electron-neutral collisions.  Secondly, it also requires including
the perpendicular component of the anisotropic electrical resistivity tensor
(the Pedersen resistivity), which is described by

\be
\eta_{\rm per} = \frac{[(1+\Gamma)^2+M_e^2]}{(1+\Gamma)\eta_{\rm par}} \\
\ee
\be
\eta_{\rm par} = \frac{M_e(f_{ei}+f_{en})}{n_e~c^2}\\
\ee
\be
\Gamma = (\frac{n_H}{n_{\rm tot}})^2~M_e~M_i \\
\ee
\be
M_e=\frac{f_{ce}}{f_{ei}+f_{en}}, ~~ M_i=\frac{f_{ci}}{f_i} , \\
\ee
where $f_i=\frac{m_H}{m_i+m_h}$.
$\eta_{\rm par}$ is the Spitzer resistivity that applies along the magnetic
field and $\eta_{\rm per}$ is the Pedersen resistivity perpendicular to it.

The left panel of Fig.~4 shows that the photospheric and
chromospheric plasma is weakly ionized but strongly magnetized for both
electrons and ions, and therefore that $\Gamma \gg 1$ for a longitudinal
photospheric magnetic field of 20~G. That suggests that the role of the Hall
effect should be negligible in the atmosphere.

The Pedersen resistivity is given as
\be
\eta_{\rm per} \propto \Gamma~\eta_{\rm par} \propto \frac{B_z^2}{n_H^2~\sqrt{T}} .
\label{ch5:eq:21}
\ee

The right panel in Fig.~4 shows that the Pedersen resistivity is
dominant throughout most of the solar chromosphere and through the entire
chromosphere of giants and supergiants.  Pedersen resistivity is $4-6$ orders
of magnitudes greater than the Spitzer resistivity, and it should therefore be
critical for the heating rates introduced by the dissipation of electric
currents that are induced by the transverse motions of Alfv\'en waves. This
result also means that the less stratified chromospheres of giants and
supergiants (i.e.~less surface gravity) will reduce the density of the
chromospheric plasma and therefore increase the significance of the Pedersen
resistivity relative to the chromospheres of dwarf stars.

This approach was recently applied to model chromospheric heating in red giants
(Airapetian et al.~2014). They employed a 1.5-D MHD code with a generalized
Ohm's law and McMurry's semi-empirical model for $\alpha$~Tau to simulate the
propagation of harmonic Alfv\'en waves at a single frequency of 0.01~mHz.  The
single fluid fully non-linear resistive and viscous MHD equations were treated
for partially-ionized plasma according to

\begin{eqnarray}
&&\frac{\partial \, \rho }{\partial \, t} +\nabla \cdot \left(\rho
\, \vec{V}\right)=0, \label{cont:eq}\\
&&\rho \left[\frac{\partial\, \vec{V}}{\partial \, t} +\left(\vec{V}\cdot \nabla \right)\,
\vec{V}\right]=-\nabla \, p + \rho \vec{g} + \vec{J}\times \vec{B}+ \nabla \vec{S} \label{mom:eq} \\
&&\frac{\partial \, \vec{B}}{\partial \, t} =-\nabla \times \vec{E},\\ &&
\vec{E}=- \vec{V}\times \vec{B} + \left( \eta_{\rm par} \vec{J}_{\rm par} + \eta_{\rm per} \vec{J}_{\rm per}\right),
\label{E:eq} \\
&&\frac{\partial \, (\rho E)}{\partial \, t}+ \nabla (\rho E \vec{V}) =
- P \nabla {\vec{V}} + \left( \eta_{\rm par} \vec{ J}_{\|}^2 + \eta_{\rm per}
\vec{J}_{\bot}^2 + \zeta_{ij} S_{ij} - L_{\rm rad} \right)
\end{eqnarray}
$S_{ij}$ denote the components of the stress tensor $\vec S = \nu
[\zeta_{ij}-(\delta_{ij} \nabla {\vec V})/3]$ and $\zeta_{ij}=\frac{1}{2}
(\frac{\partial \, V_i}{\partial \, x_j} + \frac{\partial \, V_j}{\partial \,
x_i})$; $\nu = \nu_{\rm nn} + \nu_{\rm in}$ is the viscosity coefficient due to
neutral--neutral and ion--neutral collisions (Leake et al.~2013), $E$ is the
specific internal energy, given by $E = \frac{P}{\rho(\gamma-1)}+(1-\xi_n)
\frac {X_i}{m_{av}}$, $L_{\rm rad}$ is the total radiative cooling rate; other
variables have their usual meaning.  The method of solving the continuity,
momentum and induction equations in a partially ionized plasma has been
described in detail by Arber et al.~(2001) and Leake \& Arber (2006).

A steady-state flux of Alfv\'en waves is launched from the photosphere along a
vertically diverging flux tube (Airapetian et al.~2014). The initial Poynting
flux of Alfv\'en waves is given as $F_A = 4 \times
10^7$~ergs~cm$^{-2}$~s$^{-1}$.  As waves propagate upward along the magnetic
field lines into the stellar chromosphere, the wave amplitude grows as the
density falls with height. Figure~5 shows that the amplitude of
Alfv\'en waves increases by a factor 10 at $t =0.3~t_A$, where $t_A$ is the
Alfv\'en transit time. Such an amplitude of unresolved turbulence formed by
Alfv\'en waves is consistent with the non-thermal broadening that is observed
in UV lines.

One important feature revealed by the simulations of these non-linear
transverse waves is their conversion into longitudinal (compressible) waves.
This conversion occurs at the altitude ($\geq$ 0.05 $R_{\star}$) at which the
sound speed becomes equal to, or greater than, the Alfv\'en speed (upper right
panel of Fig.~3).  At the narrow layers where plasma $\beta \sim$
1, non-linear transverse Alfv\'en wave motions become strongly coupled to
compressible wave motions or slow magneto-sonic waves (left panel of
Fig.~6).  The wave mode conversion is revealed by the formation
of non-linear density fluctuations at amplitudes as high as 50\% of the
unperturbed density, starting at 0.05 $R_{\star}$ and propagating to the upper
layers of the chromosphere as presented in the right panel of
Fig.~4. At that altitude the wave Poynting flux is about
7~$\times$~10$^5$~ergs~cm$^{-2}$~s$^{-1}$, and therefore over 70\% of the
surface energy flux is dissipated or reflected back to the chromosphere.  The
right panel of Fig.~6 shows that the total energy density reaches
its first peak at 0.05 $R_{\star}$, suggesting the formation of non-linear slow
magneto-sonic waves at an energy density of 0.25 ergs~cm$^{-3}$ in narrow
layers distributed in the chromosphere for up to 0.3 $R_{\star}$.  The energy
flux associated with compressible waves is 2
$\times$~10$^5$~ergs~cm$^{-2}$~s$^{-1}$, which is $\sim$30\% of the flux in the
Alfv\'en waves.  Highly non-linear, slow magnetosonic waves then steepen into
shocks that are not resolved in the described simulations.  The right panel
also shows that the non-linear compressible waves dissipate their energy
efficiently into heat in a narrow range of heights, from 0.05--0.4 $R_{\star}$.
The effect of the dissipation of compressible waves is traced by the sharp drop
of the compressible energy density in upward propagating waves.

The wave energy dissipates mostly via viscosity owing to neutral--neutral
collisions and resistivity caused by ion--neutral collisions.  The effect of
mode coupling of the conversion of energy from Alfv\'en waves into slow
magneto-sonic waves has also been observed and described in 1-D MHD simulations
of the solar chromosphere by Lau \& Siregar (1996), Torkelsson \& Boynton
(1998), Ofman \& Davila (1997), Airapetian et al.~(2000) and Suzuki (2013) in
simulations of solar and stellar atmospheres and winds.

At this point, the Alfv\'en wave motions become non-linear and induce
significant electric fields.  The induced perpendicular component of the
electric current (with respect to the vertical magnetic field) is then
efficiently dissipated by Pedersen resistivity with the volumetric Joule
heating rate, $\eta_{\rm per} J_{\rm per}^2$.

According to Airapetian \& Carpenter~(2013) (see also the left panel of
Fig.~3), the McMurry model predicts a strong peak in the
Alfv\'en velocity gradient that forms the reflection point at $\sim
0.2~R_{\star}$ for upward propagating Alfv\'en waves.  The model shows that
Alfv\'en waves with critical frequencies less than the critical frequency of
$\nu_{\rm crit}= dV_A/dr = 5 \times 10^{-4}$ Hz get reflected down to the
chromosphere and interact with outgoing Alfv\'en waves, thus introducing a
velocity shear.  This suggests that the chromosphere is a low-frequency filter
that passes only the higher-frequency waves into the upper atmosphere (where
they can be dissipated).  The reflected non-linear waves initiate downflows in
the chromosphere, which can potentially explain not only non-thermal broadening
but also the redshifts that are observed in UV lines from cool evolved stars.

In a partially ionized plasma, the volumetric viscous heating rate is mostly
caused by kinematic neutral--neutral viscosity in the presence of velocity
shear, $H_{\rm visc} = 0.5 \zeta_{nn} (\nabla{\delta V})^2$.  This term becomes
important in the energy balance of the chromosphere when waves become strongly
non-linear. The left panel of Fig.~7 shows that the viscous rate
becomes comparable to the Joule heating rate in the mid-chromosphere. However,
because the grid resolution is not high enough to allow us to resolve the
viscous dissipation scales, viscous effects cannot be estimated consistently.

As non-linear waves propagate upward, they exert non-linear magnetic-pressure
gradients upon the plasma, $\sim d(\vec{J} \times \vec{B})/dr$. This force is
commonly known as the ponderomotive force. It has been shown that it may be
responsible for accelerating the solar or stellar winds (see for example Ofman
\& Davila 1997, 1998; Airapetian et al. 2000, 2010).

The model by Airapetian et al.~(2014) suggests that non-linear waves can
deposit significant momentum and cause the mass loss that is consistent with
observation.  The right panel of Fig.~7 characterizes the
momentum deposition in the chromosphere in terms of the mass-loss rate,
$\dot{M} = \rho V_r r^2$ ($V_r$ is the radial velocity of the plasma, $r$ is
radial distance), throughout the atmosphere at $t=0.3~t_{A}$.  The plot shows
that at the top of the chromosphere the mass loss rate becomes constant at
about $10^{18}$ g~s$^{-1}$ $\sim 10^{-8}~{\dot M}_{\odot}$~yr$^{-1}$ .  It also
suggests that the filling factor of the open magnetic field is about 0.1$\%$.

Future models should therefore also take radiative losses into account and
calculate the profiles and fluxes of chromospheric lines formed in the model
atmosphere (as discussed in Sect.~5).  That requires knowledge
about the heating rates that balance the radiative losses.  In the 1.5-MHD
simulations presented above, Airapetian et al.~(2014) employed for the first
time the high-resolution simulations that resolve structures at scales less
than 1,500 km. The resolution of their grid allowed them to resolve resistive
dissipation scales, and thereby yielded heating rates that were physically
meaningful.

The numerical diffusivity is given by

\be D_{\rm num}= V_A\Delta^2/L, \ee where $\Delta$ is the numerical grid
spacing, and \emph{L} is the characteristic length of the physical structure.
$D_{\rm num}$ reaches its largest value then the characteristic length equals
the grid size, $V_A \Delta$.  The Pedersen diffusivity, \be D_{\rm
Pedersen}=c^2 \eta_{\rm ped}/4{\pi}, \ee where $\eta_{\rm ped}$ is given by
Eq.~(21).  The ratio of the Pedersen to the numerical
diffusivity is therefore proportional to

\be
\frac{D_{\rm Pedersen}}{D_{\rm num}} \propto \frac{B}{n_{H}^2\Delta\sqrt{T}} ~.
\label{ch5:eq:29}
\ee
Equation~29 suggests that in the regions of stronger magnetic
fields (i.e.~active regions), the ratio of physical to numerical resistivity
increases.  Indeed, Fig.~8 shows that the ratio of the Pedersen
to the numerical resistivity is mostly greater than unity throughout most of
the chromosphere of $\alpha$ Tau.  That suggests that the calculated heating
rates directly reflect the energy input into the plasma and can be used to
calculate the radiative output from the chromosphere for direct comparison with
observations.

\subsection{Momentum Deposition by Alfv\'en Waves: Driving Winds from Cool
Evolved Stars}

The general requirements for the driver of winds from cool evolved stars
include the condition for their initiation below the sonic point of the star,
namely, $R_s = {GM_{\star}}\,/\,{2 V_s^2} \sim 10-20~R_{\star}$, where $V_s$ is
the isothermal speed of sound (Holzer 1983).  A more refined condition arises
from the requirement of initiation and acceleration of winds---for example, in
non-coronal giants within the first stellar radius and its association with the
supersonically turbulent and clumpy chromosphere (Carpenter et al.~1995).
Radiation pressure on the atmospheric plasma, which is generally accepted as
the process driving massive winds from hot stars, is too small to account for
what occurs in cool giant stars.  Dust-driven effects in wind acceleration are
not important in K and early-M giants, because there is no evidence for dust
formation close to the star (Danchi et al. 1994).  The second possible
mechanisms, namely winds driven by acoustic waves, is also not capable of
producing the required mass-loss rate (see the discussion in
Sect.~3).  The most promising mechanism to date for cool evolved
stars requires MHD waves {\it both} to heat the plasma {\it and} to accelerate
the wind.

The role of MHD waves in initiating and accelerating winds from numerous types
of stars, including the Sun, solar-types and cool evolved ones, have been
studied extensively since the early 1970s (Hollweg 1973, 1978; Heinemann \&
Olbert 1980; Hartmann \& MacGregor 1980; Ofman \& Davila 1998; Airapetian et
al.~2000; Suzuki 2007; Cranmer 2008, 2009; Airapetian et al.~2010).  The
effects of 1-D non-linear Alfv\'en waves have been studied by Lau \& Siregar
(1996) and by Boynton \& Torkelsson (1996).  Moreover, 2.5-D self-consistent
non-linear treatments of Alfv\'en waves for the solar wind and accelerations
from coronal holes have also been performed (Ofman \& Davila 1997; Ofman \&
Davila 1998).  Grappin (2002) modelled a solar wind that was driven by 2.5-D
MHD Alfv\'en waves and included both open and closed magnetic
configurations. The above studies suggest that as initially small-amplitude
torsional Alfv\'en waves, which are generated at the coronal base, propagate
upward in a gravitationally-stratified atmosphere, they become non-linear and
transfer momentum efficiently to the bulk plasma flows because of the vertical
gradient of the magnetic wave pressure. Non-linear coupling of Alfv\'en waves
excite magneto-sonic waves, which can eventually steepen into longitudinal
waves that become damped by shock formation.  Suzuki \& Inutsuka (2005) and
Suzuki (2007, 2013) pursued non-linear 1.5-D MHD simulations of Alfv\'en wave
propagation from solar and stellar photospheres into the corona.  While their
model does not account for cross-field gradients, they showed that non-linear
Alfv\'en wave dissipation in the solar atmosphere can explain its
thermodynamics due to the dissipation of compressible waves produced by mode
coupling of non-linear Alfv\'en waves.

Recent progress in our understanding and modelling of coronal solar and stellar
winds driven by MHD waves provides a solid foundation to characterize such
winds in the environments of magnetically-active solar-like and evolved
luminous stars (Ofman \& Davila 1997, 1998; Cranmer 2009; Airapetian et
al.~2010; Cranmer \& Saar 2011).  If the hot, fast, tenuous winds from
solar-like stars are driven by a combination of coronal gas pressure and
Alfv\'en wave pressure, the cool massive winds emanating from cool evolved
stars are also expected to be powered by Alfv\'en wave pressure; Airapetian et
al.~(2000, 2010, 2011) describe models of slow massive winds from late-type
luminous stars.  Future models should apply a single fluid MHD approximation to
treat large-scale wind flows from magnetically open fields that extend from the
base of the wind to 25\,$R_{\star}$ and beyond.  The validity of the MHD
approximation is warranted by GHRS spectroscopic observations of wind
plasmas that have electron densities of $10^9$--10$^{10}$\,cm$^{-3}$ and
temperatures between 10,000\,K and 1\,MK.  Because the magnetic field at
the base of the wind is about 1--10\,G, the ratio of the thermal to magnetic
pressure in the plasma, or plasma-$\beta$, is less that unity at that level.

Previous models have described the propagation, in a gravitationally-stratified
atmosphere, of non-linear Alfv\'en waves that are launched from a chromospheric
hole at the base of the wind at a single frequency in two wind geometries, and
applied it to giants and to supergiants such as $\alpha$ Ori (Airapetian et
al.~2010).  The results of those simulations suggest the formation of a
two-component wind for giant stars from a magnetically open configuration that
is dubbed a `chromospheric hole'.  Figure~9 presents 2-D maps of
the radial velocities in that model. It shows the formation of slow (0.1 $V_A$)
and fast (at 0.22 $V_A$) components of the wind.  While the fast component is
formed outside the low-density region, the slow wind component emerges from the
inner regions of the low-density regions and is about 10 times less dense.

This model showed that in order to explain the high mass-loss rate of $\alpha$
Ori, a surface magnetic field of 200~G must be assumed for its
chromospherically active regions.  Magnetic fields with an average surface
distributed field of 1~G have recently been detected in this object (Petit et
al.~2013).  The finding suggests that the field can reach 200~G if the filling
factor of the field is $\sim$0.5\%.

Wind models have also been extended by adding a broad-band distribution of
Alfv\'en waves that propagate in the atmosphere of late-type giants. To model
the wind from a typical red giant (the example taken was that of $\alpha$ Tau,
K5\,III), well-constrained input model parameters were adopted (mass, radius,
temperature, initial amplitude and spectral range of Alfv\'en waves launched at
the base of the wind), together with output wind parameters inferred from
observation (terminal velocity of the wind and mass-loss rate). In order to get
a better constraint on the lowest end of the frequency spectrum of the Alfv\'en
waves ($\nu_1$), the parametric dependence of $\nu_1$ on the mass-loss rate and
wind speed were assessed by calculating four Alfv\'en wind models (Models A, B,
C, and D) corresponding to various lowest frequencies $\omega _1$ = 1, 3, 9 and
18$\tau_A^{-1}$ (or wave periods of the Alfv\'en wave spectrum of 16.2, 5.4,
1.8 and 0.9 days, respectively) (Airapetian et al.~2010).  At the outer
boundary, open (non-reflecting) conditions for MHD waves were imposed in order
to reach a quasi steady-state solution.  Airapetian et al.~(2010) solved MHD
equations for a fully ionized plasma in order to to calculate the time-averaged
terminal velocities and output mass-loss rates of quasi steady-state winds for
Models A--D.  Note that Model~D ($\omega_1/\omega_A$ = 18), corresponding to
freely propagating waves, drives faster and less massive winds.
The total mass-loss rate from a stellar atmosphere filled by an open magnetic
ﬁeld is proportional to the surface magnetic flux (Eq.~11 of Airapetian et
al.~2010).  That equation suggests that the wind mass-loss rate should vary on
timescales of the emergence and evolution of the magnetic ﬂux associated with
an open magnetic ﬁeld, or on the timescale of the rotation period of the star.  
While the timescale of magnetic-ﬁeld ampliﬁcation by giant cell convection is of the
order of 25 years (Dorch 2004), the rotation period of a typical giant is of
the order of 1 year.  

The fact that the wind forms in anisotropically distributed
`active regions' on the stellar surface also implies that the generated wind
outflows should have anisotropic and clumpy structures. These predictions are
consistent with observations of inhomogeneous chromospheres in red giants and 
supergiant stars (Eaton 2008; Ohnaka et al. 2013; Ohnaka 2013).
Observations by Mullan et al.~(1998) and Meszaros et al.~(2009) suggest that the mass-loss rate in
$\lambda$~Vel shows time variability by a factor of six, and by a factor of two in K341
(M15) and L72 (M13) over 18 months of observations.

More specifically, when the total mass-loss rate given by the model is compared
with the measured rate for $\alpha$~Tau, the area filling factor of the open
magnetic field over the stellar surface (a free model parameter) can be
constrained.  These models suggest that winds are initiated in magnetically
open structures (chromospheric holes) associated with active regions in cool
evolved stars.  As a new magnetic flux emerges to the stellar surface, outer parts of the magnetic
flux becomes open at heights where the kinetic energy of the flow is
greater than the magnetic energy. The Alfv\'en-wave flux generated at the
photosphere by convective motions (see Eq.~(9)) is proportional to the
surface magnetic flux (Airapetian et al. 2010).  The energy flux of the wind
is derived from the wave flux, and according to Eq.~(13) the mass loss by
the wind should be scaled by the cube of the surface magnetic field.

Results of that kind therefore allow us to provide the framework for sets of
physics-based models of winds for cool evolved stars.  However, while they
provide valuable insights into the dynamics of Alfv\'en-driven winds in giants,
they have not addressed the important aspect of the heating that is associated
with Alfv\'en waves.  Moreover, all multi-dimensional MHD simulations of
stellar winds performed to date have assumed a fully-ionized plasma, an
approximation that is not appropriate for the atmospheres and winds of cool
evolved giants (Suzuki 2007; Airapetian et al.~2010).

As discussed in Sect.~4.1, the effects of ion--neutral coupling
have a major effect on the propagation of Alfv\'en waves in a stellar
chromosphere. It is therefore expected that the wave momentum deposited in the
upper atmosphere should be significant in driving cool stellar winds.  This new
generation of models should be applied to larger sets of non-coronal giants and
supergiants in order to assess the generation of mass-loss and winds as a
function of stellar effective temperature, surface gravity and frequency
spectrum of MHD waves.

In conclusion, we should mention a theoretical study by Shukla \& Schlickeiser
(2003) of the effect of Alfv\'en wave propagation through a charged dusty
environment. Their model produces dust acceleration through the ponderomotive
forces exerted by non-linear waves.  It would be interesting to explore its
applicability for producing massive dusty winds in AGB stars like late-M Mira
variables. The recent detection of a surface magnetic field in a Mira star
($\chi$ Cyg) may suggest the existence of magneto-convective turbulence that is
capable of exciting MHD waves (Lobel et al.~2000; Lebre et al.~2014).

\section{Future Work: Toward Self-Consistent MHD Models of Stellar Atmospheres
and Winds}

Motivated by the recent progress in multidimensional MHD simulations of
solar and stellar chromospheres and winds from luminous late-type stars, (i.e.,
Mart{\'{\i}}nez-Sykora et al.~2012; Cheung \& Cameron 2012; Tu \& Song 2013;
Airapetian et al.~2014), we can specify four major
goals for future self-consistent atmospheric models in ways that can provide
efficient direct comparison with observations. \\

\noindent
(1) The model should resolve physically meaningful scales of energy
dissipation. A computationally efficient 1.5-D model is a potentially
useful tool to achieve a grid resolution in which the numerical
resistivity and viscosity are smaller than their physical values at
typical chromospheric parameters.  As we move from 2-D to 3-D models,
the implementation of such fine grids becomes computationally
challenging.  In general, because Spitzer resistivity is usually
extremely small, even one-dimensional models encounter problems with
the huge number of steps and small time increments. However, the
situation becomes attractive in a stellar chromosphere, where Pedersen
resistivity is over 4--6 orders of magnitude greater than the Spitzer
resistivity. Currently, 1-D and 2-D come close to resolving resistive
effects; however, the proper inclusion of the viscous dissipation is
an even more challenging task, because it requires a resolution that is
orders of magnitude more fine. The progress of implementing finer
grids will take new physics into consideration, such as the
propagation, dissipation, mode conversion and reflection of Alfv\'en
waves in partially ionized atmospheres. \\

\noindent
(2) Physically realistic chromospheric models should include the radiative
cooling rate from optically thick chromospheric environments in a
self-consistent manner. Radiative losses represent a major energy sink in a
stellar chromosphere because thermal conduction is negligibly small at
temperatures less than a few 0.1 MK.  Anderson \& Athay (1989) suggested using
an `effectively thin' approximation for a stellar chromosphere, because
emission lines of Fe~{\sc ii}, Ca~{\sc ii} and Mg~{\sc ii} are the main
contributors to radiative losses under effectively thin conditions; they
derived a simple analytical form for the total radiative loss at $T \le$
10,000\,K.  Goodman \& Judge (2012) recently generalized their expression for
total radiative loss in terms of a three-level `hydrogen' atom with two
excited states that is valid for $ T < $15,000\,K. That expression can provide
a good starting point for high-resolution numerical models.  The next step will
include the coupling between MHD and radiative transfer codes for a non-LTE
atmosphere similar to the features implemented in the 3-D radiative-MHD codes
like Bifrost and MURaM, which are used to model the solar atmosphere
(Mart{\'{i}}nez-Sykora et al.~2012). The recent version of the code includes
the generalized Ohm's law for both electrons and ions to account for the
effects of partial ionization.  However, the limitations of 3-D models
described in item 1 above do not allow the underlying code to yield physically
meaningful heating rates for comparison with observations.  For example, with
the finest resolution used in the 2-D MHD Bifrost code, the numerical
resistivity is from one to three orders of magnitude greater than the Pedersen
resistivity of Mart{\'{i}}nez-Sykora et al.~(2012).  That suggests than in the
near future 2-D and 3-D radiative-MHD models can provide a realistic
description of solar and stellar atmospheres. \\

\noindent
(3) One can generalize the chromospheric models to two and three dimensions
that apply the heating in a parameterized but physically meaningful way,
coupled to the full radiative transfer numerical code.  Once dissipation rates
are characterized in 1.5-D models, one can scale them into the energy
equation described by multidimensional models.  Such models will consistently
describe the propagation, dissipation, mode conversion and reflection of
Alfv\'en waves in partially ionized atmospheres. \\

\noindent
(4) In order to study the heating and acceleration of stellar winds
consistently with chromospheric heating simulations, future global MHD models
should be capable of extending the outer boundary to tens of stellar radii.
Future models should include a unified, fully thermodynamic model of
chromospheres of evolved luminous stars that are heated by Alfv\'en waves and
have Alfv\'en wave driven winds; models should be given for a significant range
of mass, including intermediate ones.  That new generation of models will allow
one to study the thermodynamics of winds from cool evolved stars, including
(but not limited to) the primaries of \za~systems.  Such efforts will also
allow conditions to be set for future calculations of synthetic spectra.
Important efforts will also consider the relevance of other kinds of processes,
such as magnetic and non-magnetic chromospheric turbulence and waves,
especially for low-gravity supergiants.  The latter processes have the general
ability to lift material closer to the critical point, and so helping to
support the initiation of mass loss (e.g., Schr\"oder \& Cuntz 2005; Suzuki
2007, 2013; and references therein).  Subsequent ideas and formalisms for the
treatment of magnetic and non-magnetic processes for the initiation of mass
loss in evolved coronal and non-coronal stars have been forwarded by Cranmer \&
Saar (2011) while taking stellar magnetic activity into account by extending
standard indicators of age, activity and rotation to include the evolution of
the filling factors of photospheric magnetic regions.

The X-ray luminosity-to-bolometric luminosity ratio, ($L_{\rm X}/L_{\rm bol}$, varies dramatically 
across the `dividing line' (Linsky \& Haisch 1979). Rapidly rotating early type (G-K3 type) giants on the left side of 
the `dividing line' show up to 6 orders of magnitude greater ($L_{\rm X}/L_{\rm bol}$
than the slow rotating (K3-M type) giants (Ayres 2005). This suggests that the rate of stellar rotation governs the amount of X-ray emission observed in stars across the dividing line. If the magnetic field in
  evolved stars is generated by a magnetic dynamo, then the rotation rate should
  scaled with the surface magnetic field.  Observations seems to suggest that
  the rapidly rotating early-type giants contain the signatures of
  magnetically-controlled hot coronal plasma and flare activity.  Surface
  magnetic fields have been detected directly in a number of these stars
  (Konstantinova-Antova et al.~2012).  2.5-D MHD simulations of the emergence
  of magnetic flux into a partially-ionized atmosphere performed by Leake \&
  Linton (2013) suggest that larger magnetic flux emerges faster and reaches a
  greater height in the solar atmosphere.  If applied to stellar environments,
  that model would suggest that late-type, slowly rotating giants and
  supergiant stars with weak magnetic fluxes should show signatures of magnetic
  regions that slowly (i.e., within a few years) emerge into the lower parts of
  the atmosphere and form compact, closed magnetic loop structures. Correspondingly,
  rapidly rotating early-type giants should generate strong magnetic fluxes that 
  are much more buoyant and therefore will form more extended coronal
  loop regions.  In those cases the coronal X-ray emission will be determined by the
  total volume filled by the closed magnetic loops, somewhat like those
  observed in the Sun.

In that regard, it is fundamental to appreciate that there is an intricate
interplay of different processes, operating on different scales, and which are
responsible for producing the observed phenomena and their signatures.  The
development of realistic multi-dimensional MHD-wave driven atmosphere or wind
models for cool stars is therefore both timely and appropriate.

\acknowledgments
V.~S.~A. has been supported by NASA grant NNG09EQ01C, and M.~C. by grant
HST-GO-13019.02-A.  Support for Program number HST-GO-13019.02-A was provided
by NASA through a grant from the Space Telescope Science Institute, which is
operated by the Association of Universities for Research in Astronomy,
Incorporated, under NASA contract NAS5-26555.

\clearpage

\noindent
{\bf REFERENCES:}

\bigskip
\medskip

\parindent = 0pt

\medskip\longref{
Airapetian, V.S., Ofman, L., Robinson, R.D., et al.: in Cool Stars, Stellar
 Systems,   and the Sun, Proc. of 10$^{\rm th}$ Cambridge
Workshop, ed.~by R.A. Donahue,   J.A. Bookbinder
(ASP Conf.~Ser., Vol.~154, San Francisco, 1998), p.~1569}

\medskip\longref{
Airapetian, V.S., Ofman, L., Robinson, R.D., et al.: ApJ {\bf 528}, 965 (2000)}

\medskip\longref{
Airapetian, V.S., Carpenter, K.G., Ofman, L.: ApJ {\bf 723}, 1210 (2010)}

\medskip\longref{
Airapetian, V.S., Ofman, L., Sitter, E.C., Jr., et al.: ApJ {\bf 728}, 67 (2011)}

\medskip\longref{
Airapetian, V.S. \& Carpenter, K.G.: Giants of Eclipse, AAS Topical
Conference,   BAAS {\bf 45} (2013)}

\medskip\longref{
Airapetian, V.S., Leake, J., Carpenter, K.G.: ApJ, submitted (2014)}

\medskip\longref{
Alfv\'en, H.: Nature {\bf 150}, 405 (1942)}

\medskip\longref{
An, C.-H., Suess, S.T., Moore, R.L., et al.: ApJ {\bf 350}, 309 (1990)}

\medskip\longref{
Anderson, L.S., Athay, R.G.: ApJ {\bf 346}, 1010 (1989)}

\medskip\longref{
Arber, T.D., Longbottom, A.W., Gerrard, C.L., et al.: JCP {\bf 171}, 151 (2001)}

\medskip\longref{
Auri\'ere, M., Donati, J.-F., Konstantinova-Antova, R., et al.: A\&A {\bf 516}, L2 (2010)}

\medskip\longref{
Ayres, T.R., Simon, T., Stern, R.A., et al.: ApJ {\bf 496}, 428 (1998)}

\medskip\longref{
Ayres, T.R., Brown, A., Harper, G.M.: ApJ {\bf 598}, 610 (2003)}

\medskip\longref{
Ayres, T.R.: ApJ {\bf 618}, 493 (2005)}

\medskip\longref{
Barkhudarov, M.R.: Sol. Phys. {\bf 135}, 131 (1991)}

\medskip\longref{
Braginskii, S.I.: Transport Processes in a Plasma, {\bf 1} (Consultants Bureau,
New York, 1965)}

\medskip\longref{
Boynton, C.C., Torkelsson, U.: Magnetodynamic Phenomena in the Solar
Atmosphere---\indent Prototypes of Stellar Magnetic Activity, IAU
Coll.~153, ed.~by Y. Uchida,   T. Kosugi, H.S. Hudson
(Dordrecht: Kluwer, 1996), p.~467}

\medskip\longref{
Brown, A., Deeney, B.D., Ayres, T.R., et al.: ApJS {\bf 263}, 107 (1996)}

\medskip\longref{
Buchholz, B., Ulmschneider, P., Cuntz, M.: ApJ {\bf 494}, 700 (1998)}

\medskip\longref{
Carpenter, K.G., Robinson, R.D., Wahlgren, G.M., et al.: ApJ {\bf 377}, L45 (1991)}

\medskip\longref{
Carpenter, K.G., Robinson, R.D.: ApJ {\bf 479}, 970 (1997)}

\medskip\longref{
Carpenter, K.G., Robinson, R.D., Wahlgren, G.M., et al.: ApJ {\bf 428}, 329 (1994)}

\medskip\longref{
Carpenter, K.G., Robinson, R.D., Judge, P.G.: ApJ {\bf 444}, 424 (1995)}

\medskip\longref{
Carpenter, K.G., Airapetian, V.S.: Cool Stars, Stellar Systems, and the Sun,
ed.~by   E. Stempels (AIP Conf.~Series {\bf 1094}, 2009), p.~712}

\medskip\longref{
Charbonneau, P., MacGregor, K.B.: ApJ {\bf 454}, 901 (1995)}

\medskip\longref{
Cheung, M.C.M., Cameron, R.H.: ApJ {\bf 750}, 6 (2012)}

\medskip\longref{
Chiavassa, A., Haubois, X., Young, J.S., et al.: A\&A {\bf 515} A12 (2010)}

\medskip\longref{
Cranmer, S.R.: ApJ {\bf 689}, 316 (2008)}

\medskip\longref{
Cranmer, S.R.: ApJ {\bf 706}, 824 (2009)}

\medskip\longref{
Cranmer, S.R., Ballegooigen, A.A.: ApJS {\bf 156}, 265 (2005)}

\medskip\longref{
Cranmer, S.R., Saar, S.H.: ApJ {\bf 741}, 54 (2011)}

\medskip\longref{
Cuntz, M.: ApJ {\bf 353}, 255 (1990)}

\medskip\longref{
Cuntz, M.: A\&A {\bf 325}, 709 (1997)}

\medskip\longref{
Cuntz, M., Rammacher, W., Ulmschneider, P.: ApJ {\bf 432}, 690 (1994)}

\medskip\longref{
Cuntz, M., Ulmschneider, P., Musielak, Z.E.: ApJ {\bf 493}, L117 (1998)}

\medskip\longref{
Cuntz, M., Rammacher, W., Ulmschneider, P., et al.: ApJ {\bf 522}, 1053 (1999)}

\medskip\longref{
Cuntz, M., Harper, G.M., Bennett, P.D.: A\&A {\bf 376}, 154 (2001)}

\medskip\longref{
Danchi, W.C., Bester, R.W., Degiacomi, C.G., et al.: ApJ {\bf 107}, 1469 (1994)}

\medskip\longref{
Davila, J.M.: ApJ {\bf 317}, 514 (1987)}

\medskip\longref{
De Pontieu, B., Haerendel, G.: A\&A {\bf 338}, 729 (1998)}

\medskip\longref{
De Pontieu, B., Martens, P.C.H., Hudson, H.S.: ApJ {\bf 558}, 859 (2001)}

\medskip\longref{
De Pontieu, B., McIntosh, S.W., Carlsson, M.: Science {\bf 318}, 1574 (2007)}

\medskip\longref{
Doyle, J.G., Houdebine, E.R., Mathioudakis, M., et al.: A\&A {\bf 285}, 233 (1994)}

\medskip\longref{
Dupree, A.K., Lobel, A., Young, P.R., et al.: ApJ {\bf 622}, 629 (2005)}

\medskip\longref{
Eaton, J.A.: AJ {\bf 136}, 1964 (2008)}

\medskip\longref{
Eaton, J.A., Henry, G.W., Odell, A.P.: ApJ {\bf 679}, 1490 (2008)}

\medskip\longref{
Eriksson, K., Linsky, J.L., Simon, T.: ApJ {\bf 272}, 665 (1983)}

\medskip\longref{
Fawzy, D., Ulmschneider, P., St{\c e}pie{\'n}, K., et al.: A\&A {\bf 386}, 983 (2002)}

\medskip\longref{
Fontenla, J.M., Avrett, E.H., Loeser, R.: ApJ {\bf 355}, 700 (1990)}

\medskip\longref{
Fontenla, J.M., Avrett, E.H., Loeser, R.: ApJ {\bf 572}, 636 (2002)}

\medskip\longref{
Freytag, B., Steffen, M., Dorch, B.: AN {\bf 323}, 213 (2002)}

\medskip\longref{
Gary, G.A.: Sol. Phys. {\bf 203}, 71 (2001)}

\medskip\longref{
Goodman, M.L.: ApJ {\bf 533}, 501 (2000)}

\medskip\longref{
Goodman, M.L.: A\&A {\bf 424}, 691 (2004)}

\medskip\longref{
Goodman, M.L., Judge, P.G.: ApJ {\bf 751}, 75 (2012)}

\medskip\longref{
Grappin, R., L\'eorat, J., Habbal, S.R. JGR {\bf 107}, 1380 (2002)}

\medskip\longref{
Gray, D.F.: AJ {\bf 135}, 1450 (2008)}

\medskip\longref{
Guerreiro, N., Hansteen, V., De Pontieu, B.: ApJ {\bf 769}, 47 (2013)}

\medskip\longref{
Hansteen, V.: ApJ {\bf 402}, 741 (1993)}

\medskip\longref{
Hansteen, V.H., Hara, H., De Pontieu, B., et al.: ApJ {\bf 718}, 1070 (2010)}

\medskip\longref{
Hartmann, L., Avrett, E.H.: ApJ {\bf 284}, 238 (1984)}

\medskip\longref{
Hartmann, L., MacGregor, K.B.: ApJ {\bf 242}, 260 (1980)}

\medskip\longref{
Harper, G.M., Brown, A., Bennett, P.D., et al.: ApJ {\bf 129}, 1018 (2005)}

\medskip\longref{
Harper, G.M.: ApJ {\bf 720}, 1767 (2010)}

\medskip\longref{
Harper, G.M., O{'}Riain, N., Ayres, T.R.: MNRAS {\bf 428}, 2064 (2013)}

\medskip\longref{
Heinemann, M., Olbert, S.: JGR {\bf 85}, 1311 (1980)}

\medskip\longref{
Hollweg, J.V.: JGR {\bf 78}, 3643 (1973)}

\medskip\longref{
Hollweg, J.V.: Sol. Phys. {\bf 56}, 305 (1978)}

\medskip\longref{
Holzer, T.E., Fl{\aa}, T., Leer, E.: ApJ {\bf 275}, 808 (1983)}

\medskip\longref{
Holzer, T.E. \& MacGregor, K.B.: in Mass Loss from Red Giants, ed. by
M. Morris, B. Zuckerman (ASSL {\bf 117}, Dordrecht: Reidel,
1985), p.~229}

\medskip\longref{
Holzer, T.E.: in Circumstellar Matter; Proc.~IAUS 122, ed.~by
I. Appenzeller, C. Jordan (Dordrecht: Reidel, 1987), p.~289}

\medskip\longref{
Jess, D.B., Mathioudakis, M., Erd$\acute{\rm e}$lyi, R., et al.: Science
{\bf 323}, 1582 (2009)}

\medskip\longref{
Jackson, J.D.: Classical Electrodynamics, 3rd ed. (New York: Wiley 1999)}

\medskip\longref{
Judge, P.G.; Tritschler, A.; Chye Low, B.: ApJ Lett. {\bf 730}, 4, (2011)}

\medskip\longref{
Judge, P.G., Carpenter, K.G.: ApJ {\bf 494}, 828 (1998)}

\medskip\longref{
Judge, P.G., Cuntz, M.: ApJ {\bf 409}, 776 (1993)}

\medskip\longref{
Judge, P.G., Stencel, R.E.: ApJ {\bf 371}, 357 (1991)}

\medskip\longref{
Khodachenko, M.L., Arber, T.D., Rucker, H.O., et al.: A\&A {\bf 442}, 103 (2004)}

\medskip\longref{
Konstantinova-Antova, R.K., Auri\'ere, M., Charbonnel, C., et al.: A\&A
   {\bf 524}, 57 (2010)}

\medskip\longref{
Konstantinova-Antova, R.K., Auri\'ere, M., Petit, P., et al.: A\&A {\bf 541},
   44 (2012)}

\medskip\longref{
Kopp, R.A., Holzer, T.E.: Sol. Phys. {\bf 49}, 43 (1976)}

\medskip\longref{
Krasnoselskikh, V., Vekstein, G., Hudson, H.S., et al.: ApJ {\bf 724}, 1542 (2010)}

\medskip\longref{
Lau, Y.-T., Siregar, E.: ApJ {\bf 465}, 451 (1996)}

\medskip\longref{
Leake, J.E., Arber, T.D.: A\&A {\bf 450}, 805 (2006)}

\medskip\longref{
Leake, J.E., Arber, T.D., Khodachenko, M.: A\&A {\bf 442}, 1091 (2005)}

\medskip\longref{
Leake, J.E., Linton, M.G.: ApJ {\bf 764}, 54 (2013)}

\medskip\longref{
Leake, J.E., Lukin, V.S., Linton, M.G.: Physics of Plasmas {\bf 20}, 061202 (2013)}

\medskip\longref{
L\'ebre, A., Auri\'ere, M., Fabas, N., et al.: A\&A {\bf 561}, A85 (2014)}

\medskip\longref{
Linsky, J.L., Ayres, T.R.: ApJ {\bf 220}, 619 (1978)}

\medskip\longref{
Linsky, J.L., Haisch, B.M.: ApJ {\bf 229}, L27 (1979)}

\medskip\longref{
Lobel, A., Bagnulo, S., Doyle, J.G., et al.: MNRAS {\bf 317}, 391 (2000)}

\medskip\longref{
MacGregor, K.B., Charbonneau, P.: ApJ {\bf 430}, 387 (1994)}

\medskip\longref{
Mart{\'{\i}}nez-Sykora, J., De Pontieu, B., Hansteen, V.: ApJ {\bf 753}, 161 (2012)}

\medskip\longref{
Mathioudakis, M., Jess, D.B., Erd\'elyi, R.: Sp. Sci. Rev. {\bf 175}, 1 (2013)}

\medskip\longref{
Matthaeus, W.H., Zank, G.P., Oughton, S., et al.: ApJ {\bf 523},  L93 (1999)}

\medskip\longref{
McMurry, A.D.: MNRAS {\bf 302}, 37 (1999)}

\medskip\longref{
Mitchner, M. \& Kruger, C. H. Partially Ionized Gases (New York: Wiley) (1973)}

\medskip\longref{
Morton, R.J., Verth, G., Jess, D.B., et al.: Nature Comm., {\bf 3}, 1315 (2012)}

\medskip\longref{
Morton, R.J., Verth, G., Fedun, V., et al.: ApJ {\bf 768}, 11, (2013)}

\medskip\longref{
Musielak, Z.E., Ulmschneider, P.: A\&A {\bf 386}, 606 (2002)}

\medskip\longref{
Narain, U., Ulmschneider, P.: Sp. Sci. Rev. {\bf 54}, 377 (1990)}

\medskip\longref{
Narain, U., Ulmschneider, P.: Sp. Sci. Rev. {\bf 75}, 453 (1996)}

\medskip\longref{
Neilson, H.R., Lester, J.B., Haubois, X.: in 9th Pacific Rim Conference on
Stellar   Astrophysics, ed.~by S.~Qain, K.~Leung, L.~Zhu, S.~Kwok
(ASP Conf.~Ser.~{\bf 451}, San Francisco, 2011), p.~117}

\medskip\longref{
Noyes, R.W, Hartmann, L.W., Baliunas, S.L., et al.: ApJ {\bf 279}, 763 (1984)}

\medskip\longref{
Ofman, L., Davila, J.M.: ApJ {\bf 476}, 357 (1997)}

\medskip\longref{
Ofman, L., Davila, J.M.: JGR {\bf 103}, 23677 (1998)}

\medskip\longref{
Ohnaka, K.: A\&A {\bf 553}, 8 (2013)}

\medskip\longref{
Ohnaka, K., Hofmann, K.H., Schertl, D., et al.:  A\&A {\bf 555}, 18, (2013)}

\medskip\longref{
Osterbrock, D.E.: ApJ {\bf 134}, 347 (1961)}

\medskip\longref{
Pagano, I., Linsky, J.L., Valenti, J., et al.: A\&A {\bf 331} (2004)}

\medskip\longref{
P{\'e}rez Mart{\'{\i}}nez, M.I., Schr\"oder, K.-P., Cuntz, M.: MNRAS
{\bf 414}, 418 (2011)}

\medskip\longref{
Peter, H.: A\&A {\bf 449}, 759 (2006)}

\medskip\longref{
Petit, P., Auri\'ere, M., Konstantinova-Antova, R., et al.:  Lect. Notes Phys.
{\bf 857}, 231 (2013)}

\medskip\longref{
Piddington, J.H.: MNRAS {\bf 116}, 314 (1956)}

\medskip\longref{
Pijpers, F.P., Hearn, A.G.: A\&A {\bf 209}, 198 (1989)}

\medskip\longref{
Rabin, D.: ApJ {\bf 391}, 832 (1992)}

\medskip\longref{
Reimers, D., H\"unsch, M., Schmitt, J.H.M.M., et al.: A\&A {\bf 310}, 813 (1996)}

\medskip\longref{
Robinson, R.D., Carpenter, K.G., Brown, A.: ApJ {\bf 503}, 396 (1998)}

\medskip\longref{
Rosner, R., An, C.-H., Musielak, Z.E., et al.: ApJ {\bf 372}, L91 (1991)}

\medskip\longref{
Rosner, R., Musielak, Z.E., Cattaneo, F., et al.: ApJ {\bf 442}, L25 (1995)}

\medskip\longref{
Ruderman, M.S., Berghmans, D., Goossense, M., et al.: A\&A {\bf 320}, 305 (1997)}

\medskip\longref{
Rutten, R.G.M., Schrijver, C.J., Lemmens, A.F.P., et al.: A\&A {\bf 252}, 203 (1991)}

\medskip\longref{
Schrijver, C.J.: A\&A {\bf 172}, 111 (1987)}

\medskip\longref{
Schrijver, C.J., in Inside the Stars, Proc. IAU Coll. 137, ed.~by W.W.~Weiss,
    A.~Baglin (ASP Conf.~Ser, {\bf 40}, San Francisco, 1993), p.~591}

\medskip\longref{
Schrijver, C.J.: A\&A Rev. {\bf 6}, 181 (1995)}

\medskip\longref{
Schrijver, C.J., Pols, O.R.: A\&A {\bf 278}, 51 (1993)}

\medskip\longref{
Schr\"oder, K.-P., Cuntz, M.: ApJ {\bf 630}, L73 (2005)}

\medskip\longref{
Simon, T., Herbig, G., Boesgaard, A.M.: ApJ {\bf 293}, 551 (1985)}

\medskip\longref{
Skumanich, A.: ApJ {\bf 171}, 565 (1972)}

\medskip\longref{
Soler, R., Carbonell, M., Ballester, J.L., et al.: ApJ {\bf 767}, 171 (2013)}

\medskip\longref{
Song, P., Vasyliunas, V.M.: JGR {\bf 116}, A09104 (2011)}

\medskip\longref{
Steiner, O.: in Kodai School on Solar Physics, AIP Conf. Proc. 919, p. 74 (2007)}

\medskip\longref{
Strassmeier, K.G., Handler, G., Paunzen, E., et al.: A\&A {\bf 281}, 855 (1994)}

\medskip\longref{
Shukla, P.K., Schlickeiser, R.: Phys. of Plasmas {\bf 10}, 1523 (2003)}

\medskip\longref{
Suzuki, T.K., Inutsuka, S.: ApJ {\bf 632}, L49 (2005)}

\medskip\longref{
Suzuki, T.K.: ApJ {\bf 659}, 1592 (2007)}

\medskip\longref{
Suzuki, T.K.: AN {\bf 334}, 81 (2013)}

\medskip\longref{
Tomczyk, S., McIntosh, S.W., Keil, S.L., et al.: Science {\bf 317}, 1192 (2007)}

\medskip\longref{
Torkelsson, U. \& Boynton, G.C.: MNRAS {\bf 295}, 55 (1998)}

\medskip\longref{
Tu, J., Song, P.: ApJ {\bf 777}, 53 (2013)}

\medskip\longref{
Ulmschneider, P.: A\&A {\bf 222}, 171 (1989)}

\medskip\longref{
Ulmschneider, P., Musielak, Z.E., Fawzy, D.E.: A\&A {\bf 374}, 662 (2001)}

\medskip\longref{
Vernazza, J.E., Avrett, E.H., Loeser, R.: ApJS {\bf 30}, 1   (1976)}

\medskip\longref{
Vernazza, J.E., Avrett, E.H., Loeser, R.: ApJS {\bf 45}, 635 (1981)}

\medskip\longref{
Velli, M.: A\&A {\bf 270}, 304 (1993)}

\medskip\longref{
Wood, B.E., Linsky, J.L., Ayres, T.R.: ApJ {\bf 478}, 745 (1997)}

\clearpage

\begin{figure*}
\centering
\begin{tabular}{cc}
\epsfig{file=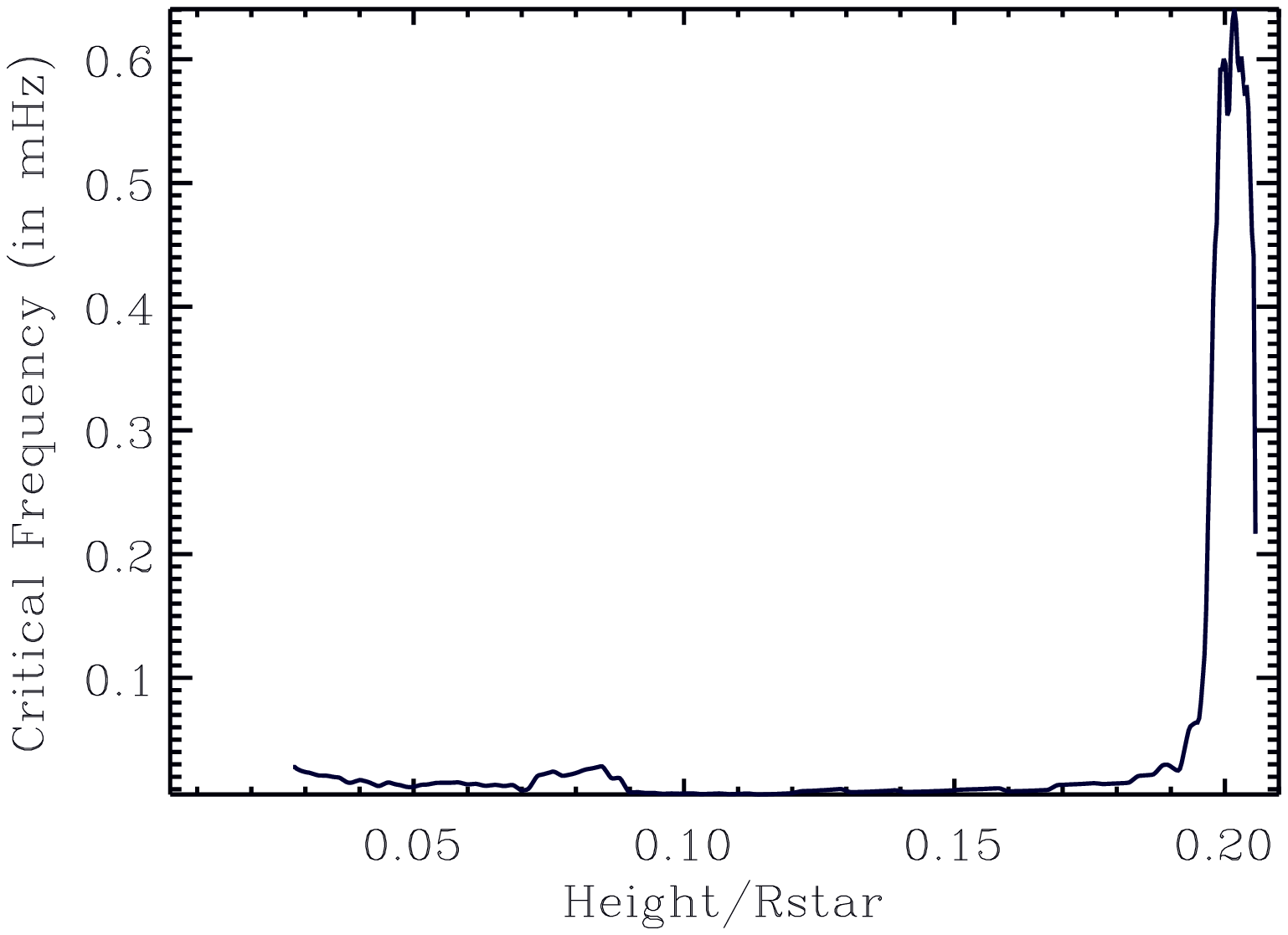,width=0.45\linewidth} &
\epsfig{file=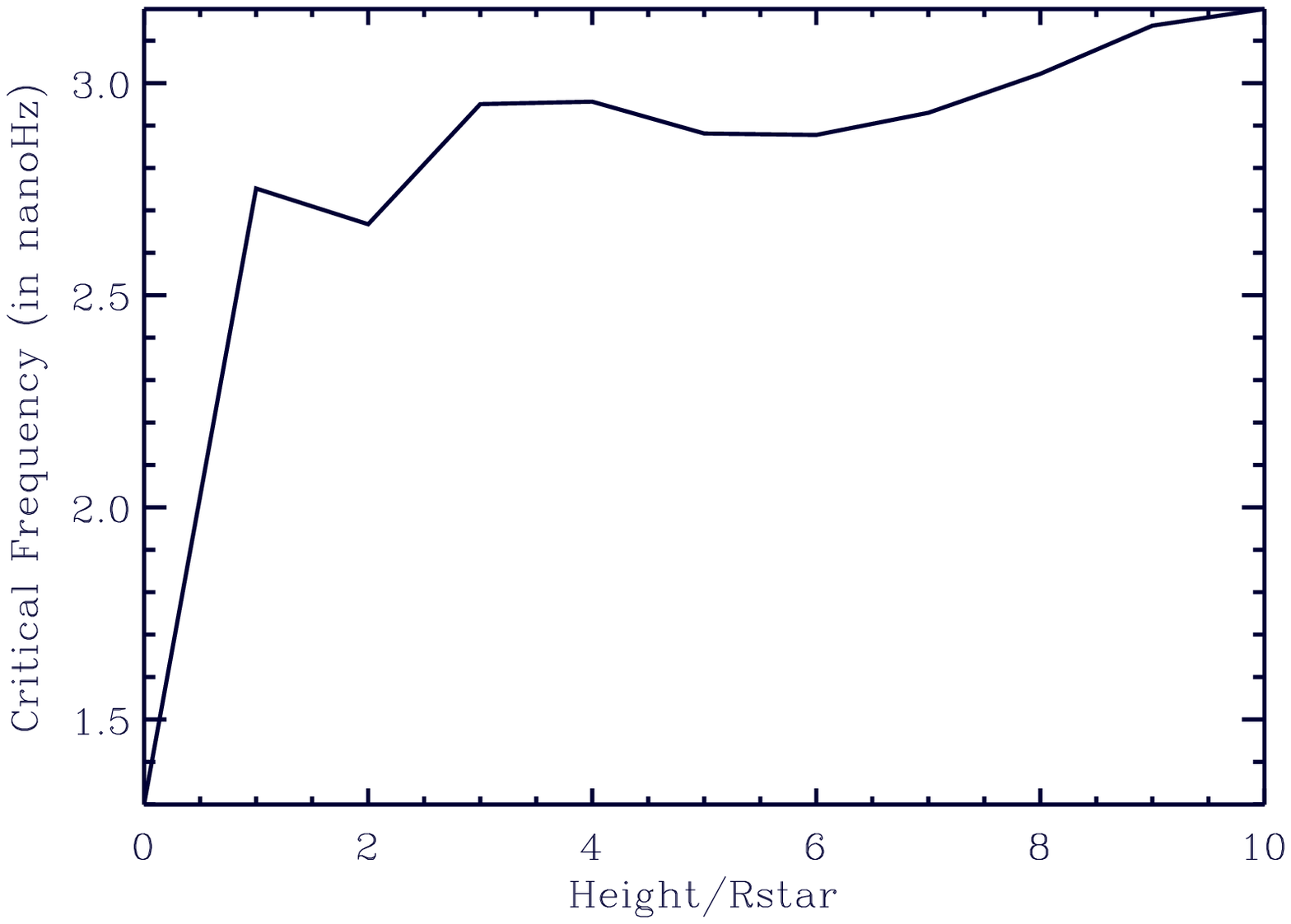,width=0.45\linewidth}
\end{tabular}
\caption{
Radial profiles of the critical frequency in a solar or stellar
chromosphere as predicted by semi-empirical models of $\alpha$ Tau
({\it left}\,) and 31 Cyg ({\it right\,}).
}
\end{figure*}

\clearpage

\begin{figure*}
\centering
\begin{tabular}{c}
\epsfig{file=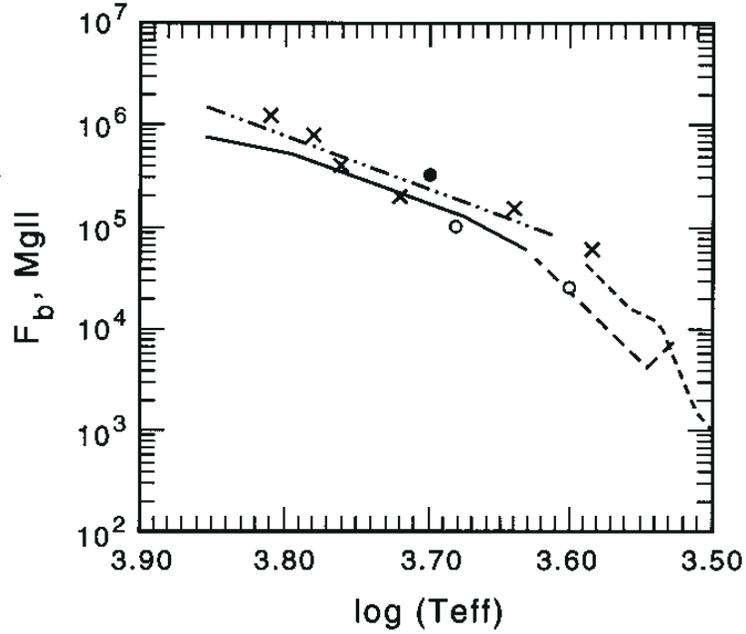,width=0.65\linewidth}
\end{tabular}
\caption{
The basal flux limit for Mg~{\sc ii} {\it h}+{\it k} is shown as a
function of effective temperature and compared to results from observations.
The solid line represents the results of Rutten et al.~(1991), and the
double-dotted/dashed line the earlier ones from Schrijver (1987).  The
long-dashed line shows the upper limits derived for very cool dwarfs (Doyle et
al.~1994).  The short-dashed line is the Mg~{\sc ii} basal flux limit for
giants (Judge \& Stencel 1991).  The crosses and open circles represent
theoretical results from the present work, for main-sequence stars and giants,
respectively.  The closed circle at log~$T_{\rm eff}$ = 3.70 is the theoretical
result of Cuntz et al.~(1994) but ignores the small differences in the Mg~{\sc
ii} flux due to the different atmospheric metallicities considered (modified
version of figure~5 of Schrijver 1995).  (From Buchholz et al.~1998.)
}
\end{figure*}

\clearpage

\begin{figure*}
\centering
\begin{tabular}{cc}
\epsfig{file=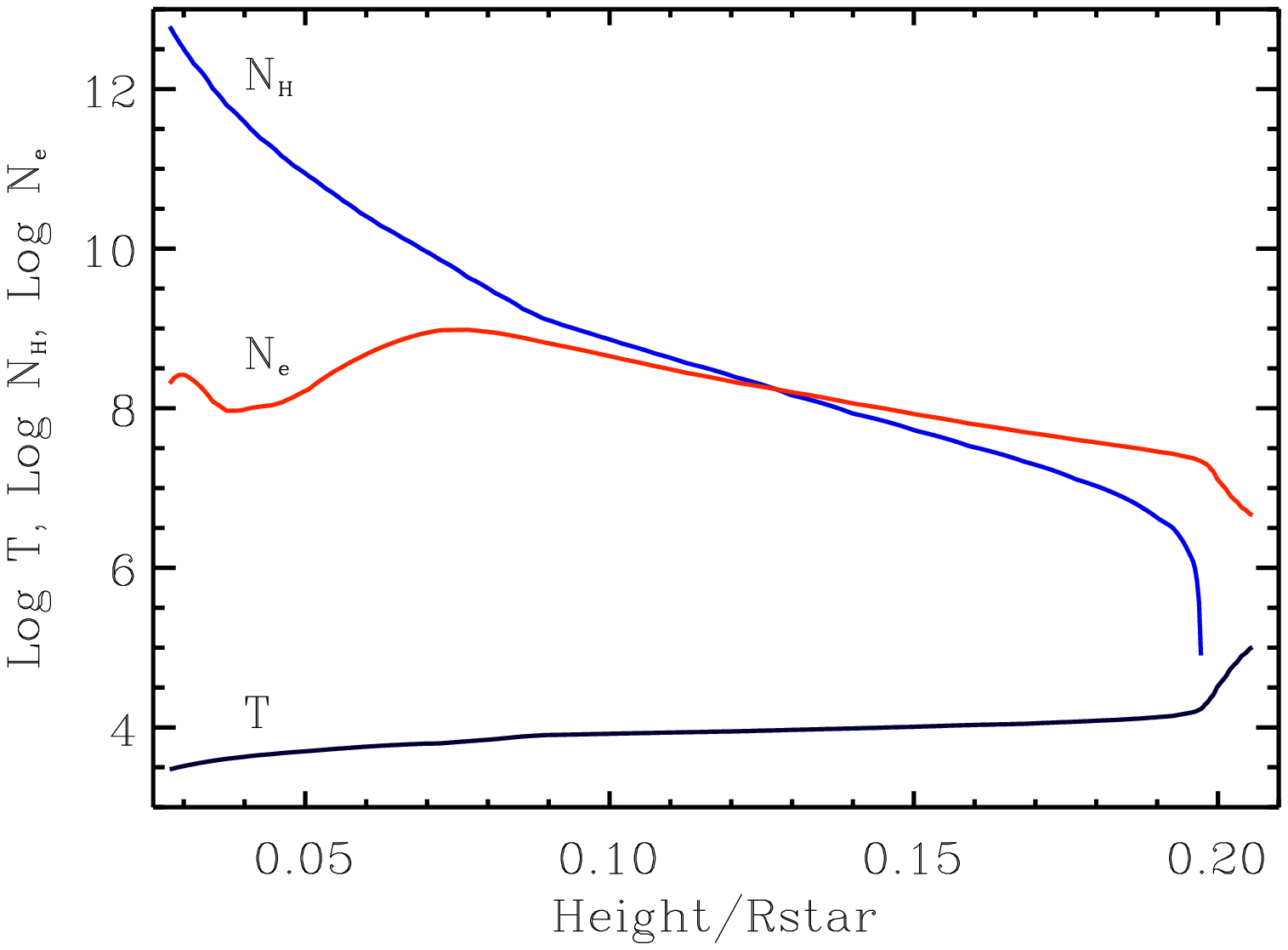,width=0.45\linewidth} &
\epsfig{file=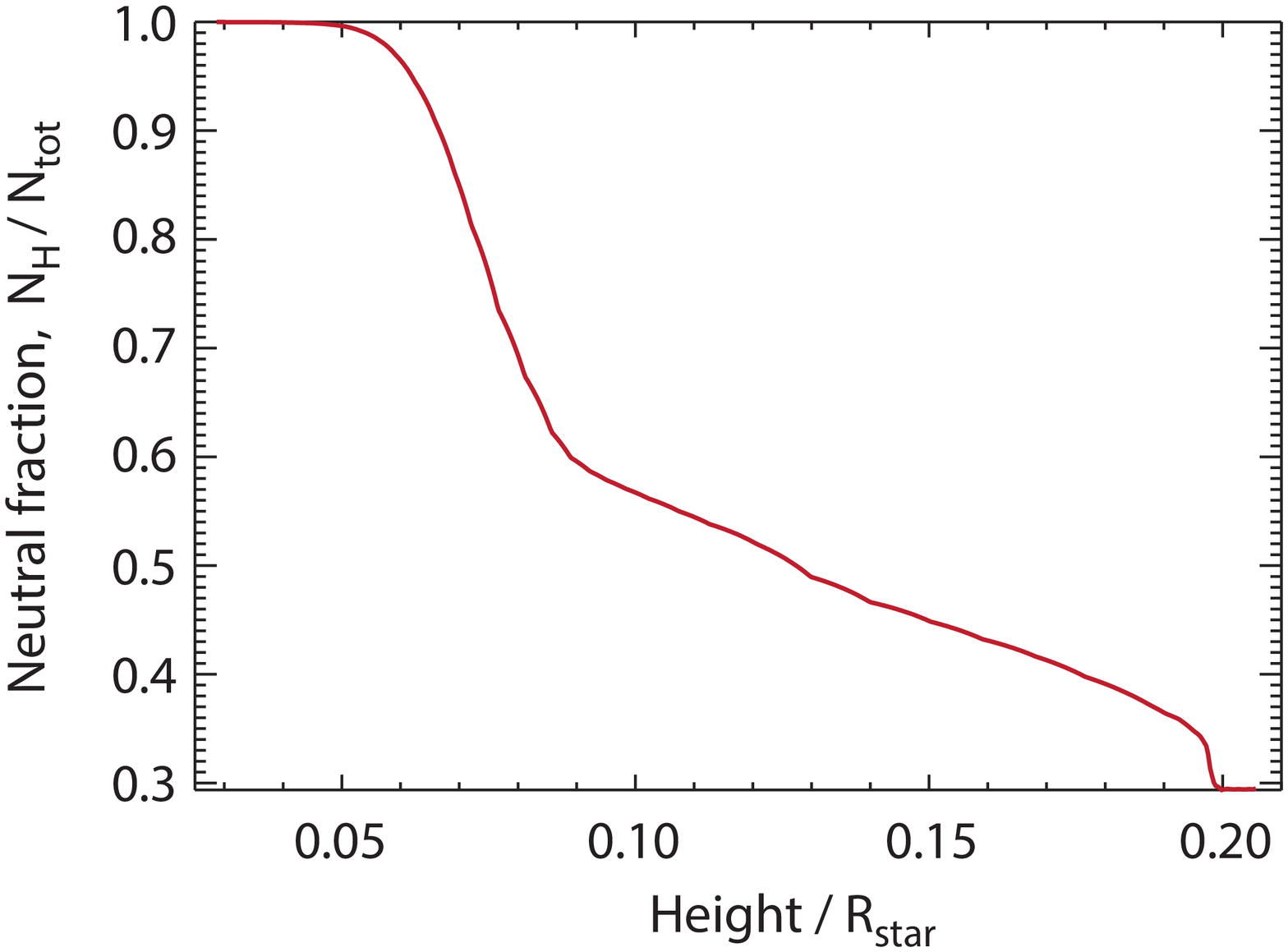,width=0.45\linewidth}
\end{tabular}
\caption{
Chromospheric model of $\alpha$ Tau. {\it Left:} radial profiles
of $T$, $N_H$, $N_e$.  {\it Right:} radial profile of neutral fraction
$N_H/N_{\rm tot}$.
}
\end{figure*}

\clearpage

\begin{figure*}
\centering
\begin{tabular}{cc}
\epsfig{file=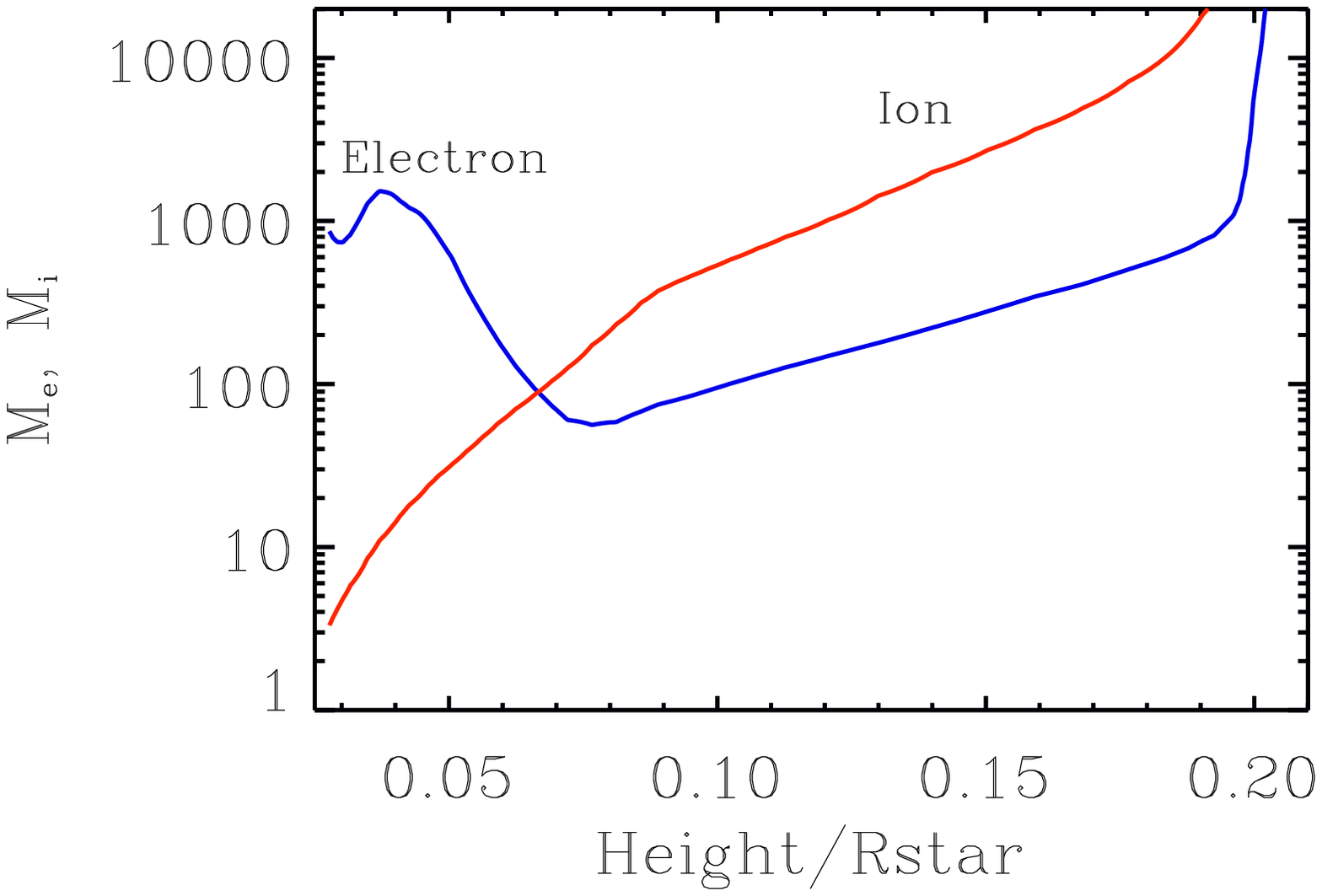,width=0.45\linewidth} &
\epsfig{file=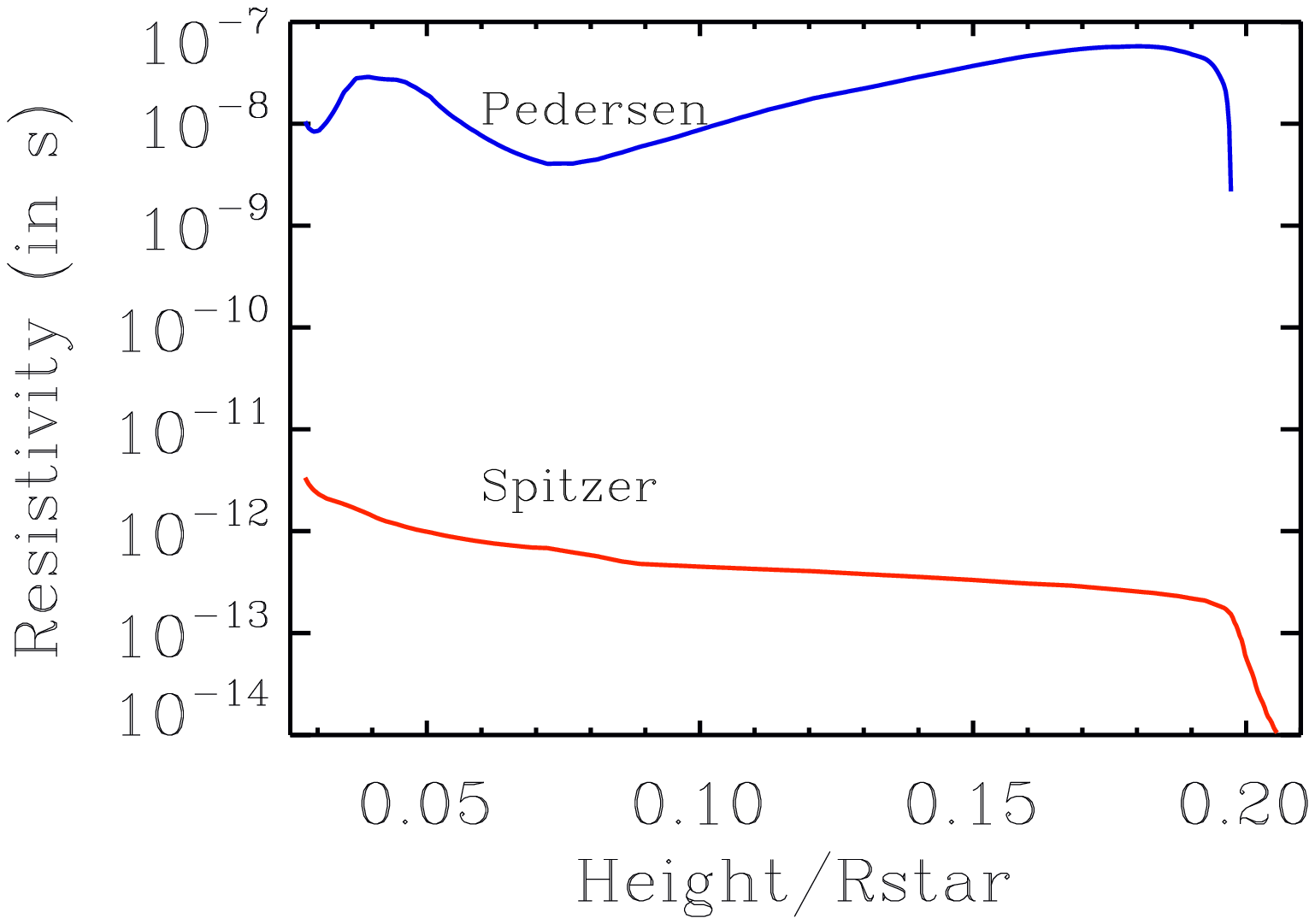,width=0.45\linewidth}
\end{tabular}
\caption{
Chromospheric model of $\alpha$ Tau. {\it Left:} radial profiles of
electron and proton magnetizations throughout the chromosphere.  {\it Right:}
Spitzer and Pedersen resistivity throughout the chromosphere.
}
\end{figure*}

\clearpage

\begin{figure*}
\centering
\begin{tabular}{c}
\epsfig{file=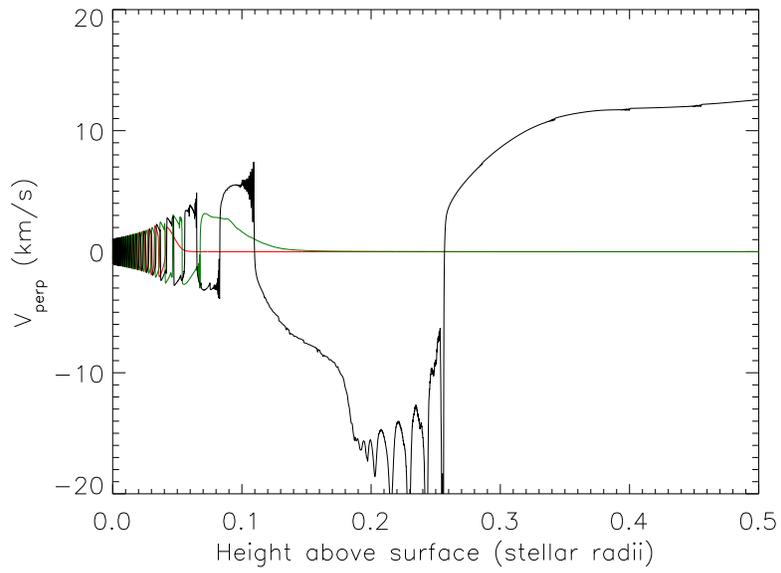,width=0.65\linewidth}
\end{tabular}
\caption{
Model outputs: radial profile of the Alfv\'en wave amplitude at
$0.1~t_A$ (red), $0.2~t_A$ (green) and $0.3~t_A$ (black).
}
\end{figure*}

\clearpage

\begin{figure*}
\centering
\begin{tabular}{cc}
\epsfig{file=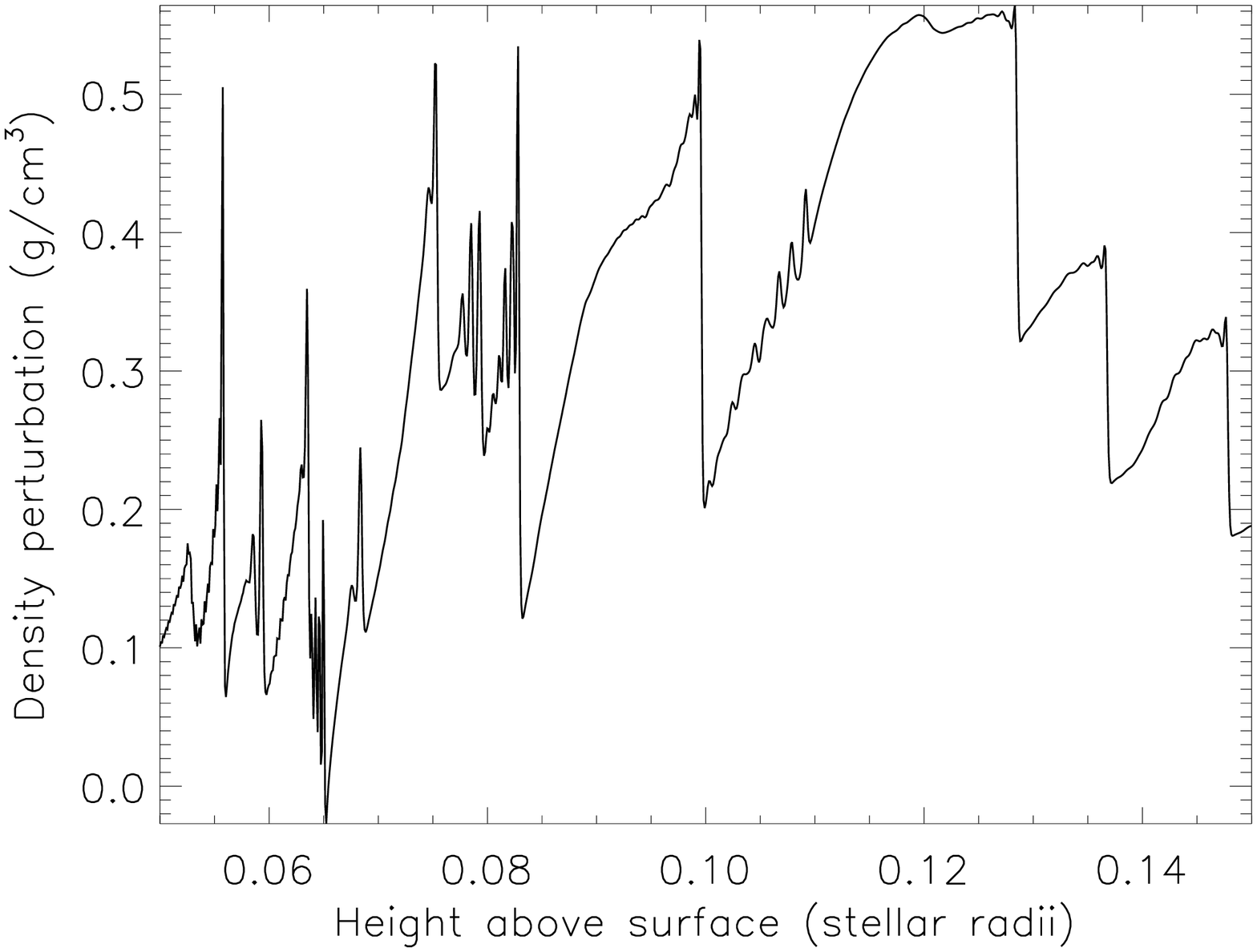,width=0.45\linewidth} &
\epsfig{file=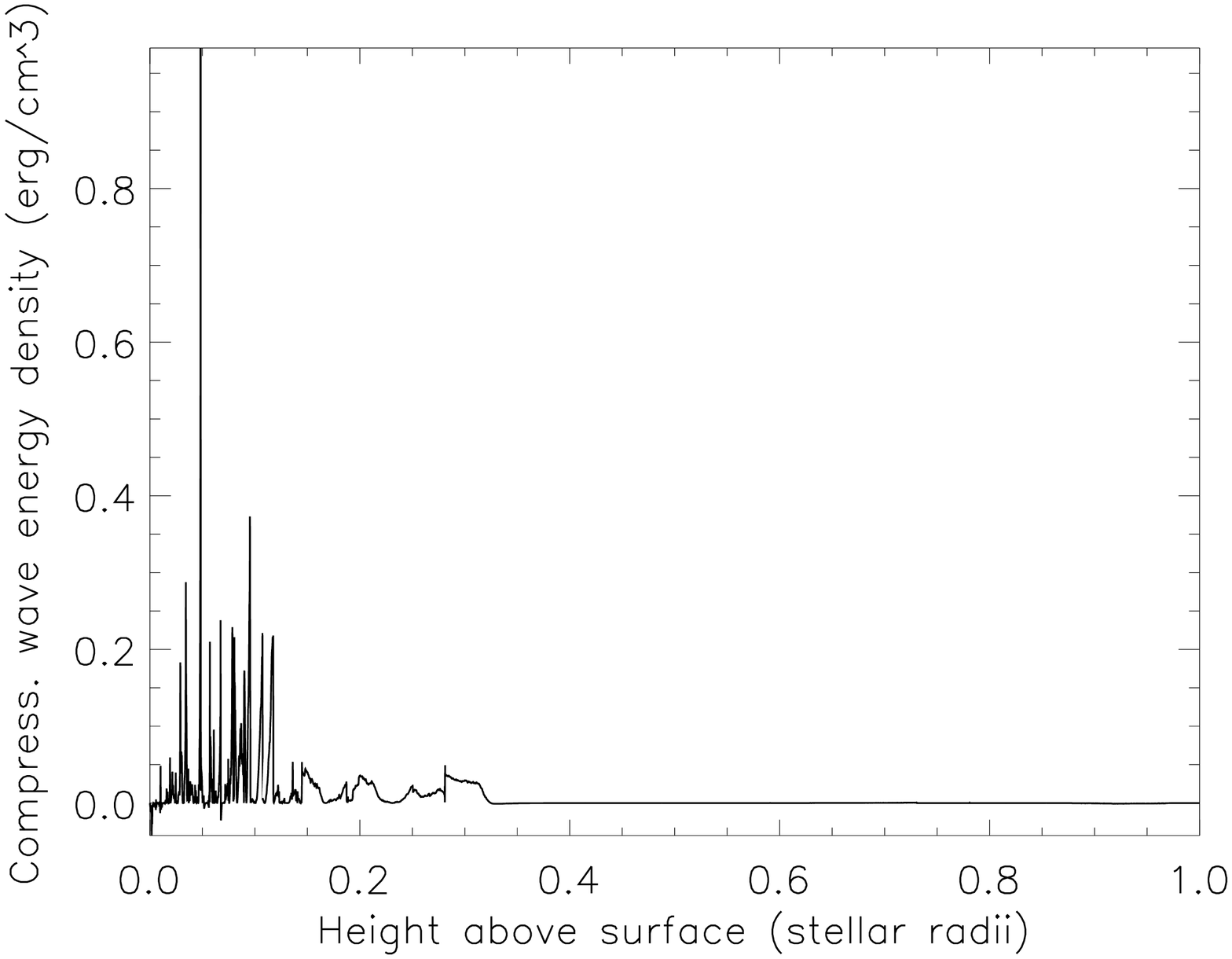,width=0.45\linewidth}
\end{tabular}
\caption{
Model outputs. {\it Left:} radial profile of the density perturbation,
normalized to the background plasma density, shows the presence of non-linear
slow magneto-sonic waves at $t=0.3~t_A$. {\it Right:} distribution of energy in
slow magneto-sonic waves concentrated in narrow layers throughout the
chromosphere at $t=0.3~t_A$.
}
\end{figure*}

\clearpage

\begin{figure*}
\centering
\begin{tabular}{cc}
\epsfig{file=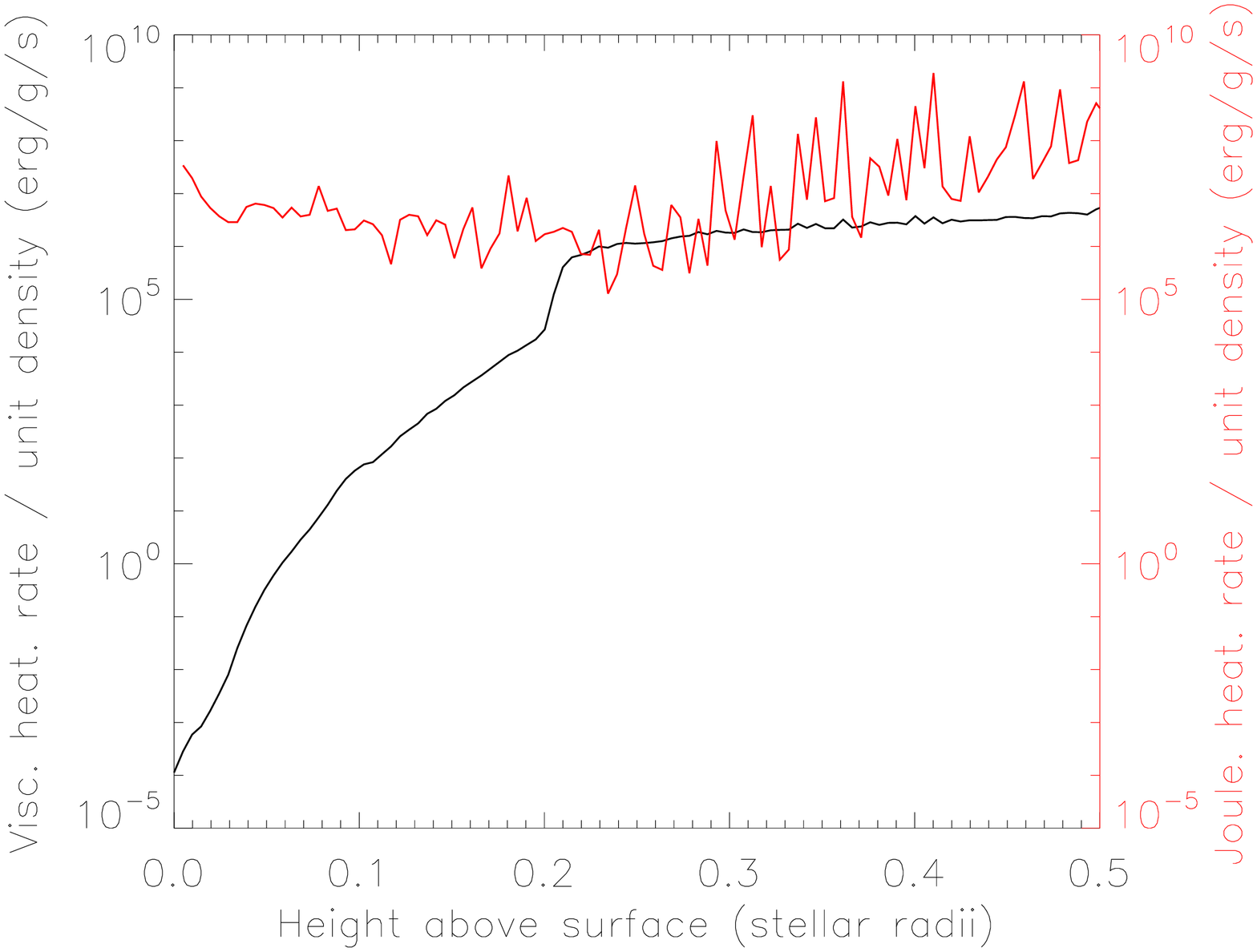,width=0.45\linewidth} &
\epsfig{file=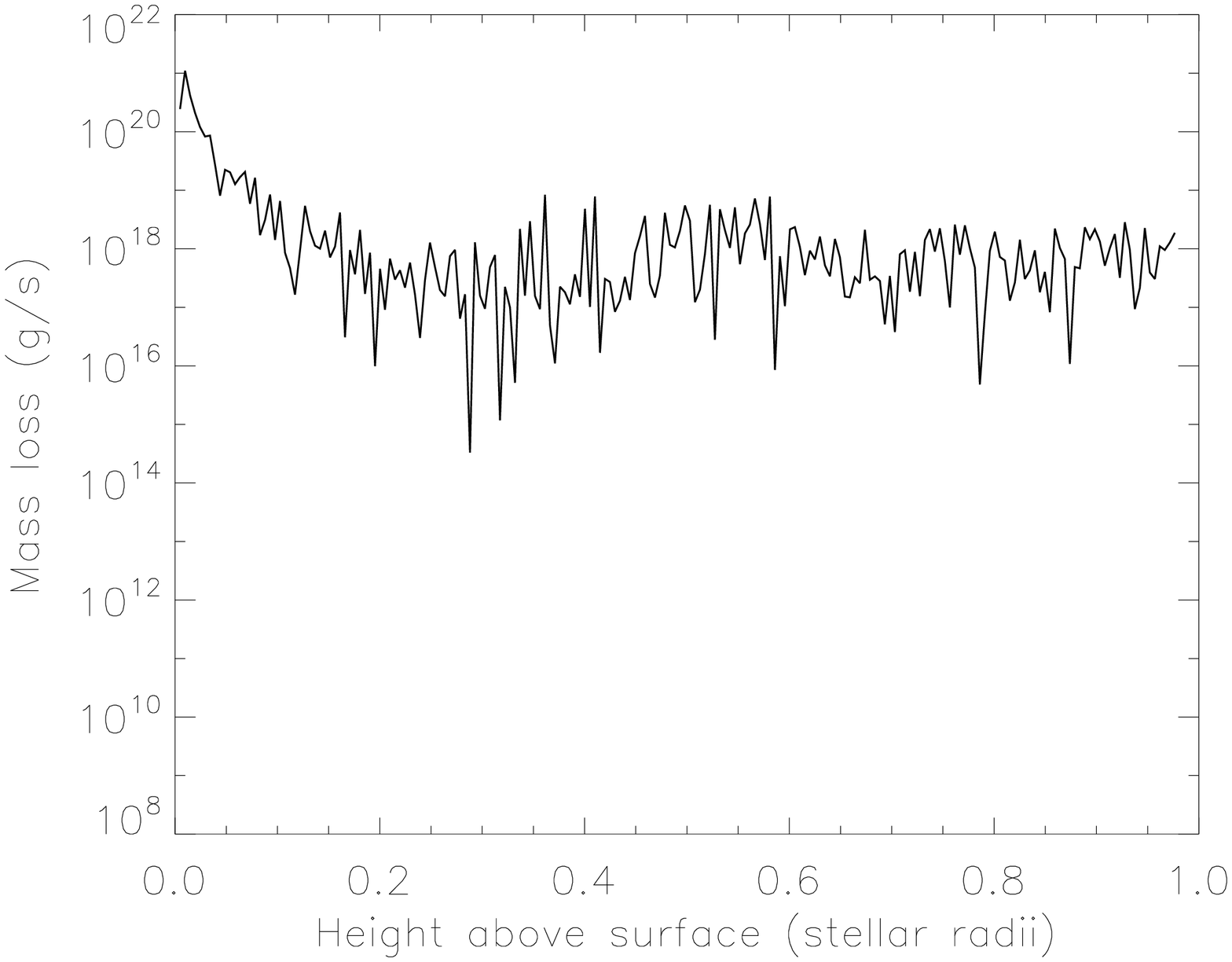,width=0.45\linewidth}
\end{tabular}
\caption{
Model outputs. {\it Left:} viscous (black) vs Joule (red) heating in
the chromosphere.  {\it Right:} radial profile of the mass-loss rate at
0.3~$t_A$.
}
\end{figure*}

\clearpage

\begin{figure*}
\centering
\begin{tabular}{c}
\epsfig{file=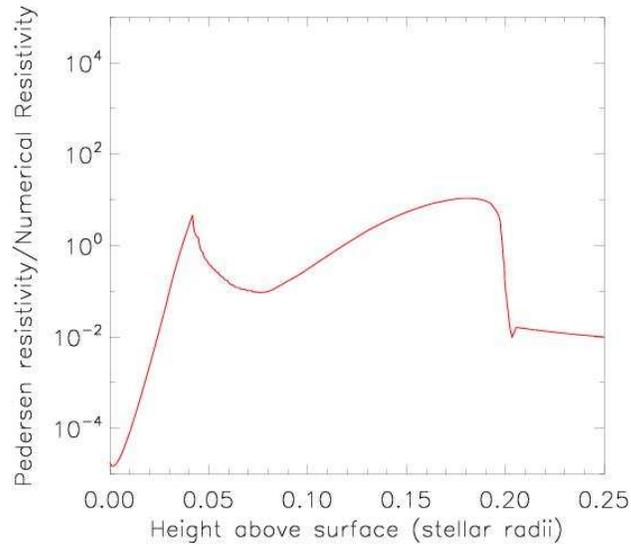,width=0.65\linewidth}
\end{tabular}
\caption{
Model outputs. The vertical profile of the ratio of the Pedersen to
the numerical resistivity in a 1.5-D MHD model by Airapetian et al.~(2014).
}
\end{figure*}

\clearpage

\begin{figure*}
\centering
\begin{tabular}{c}
\epsfig{file=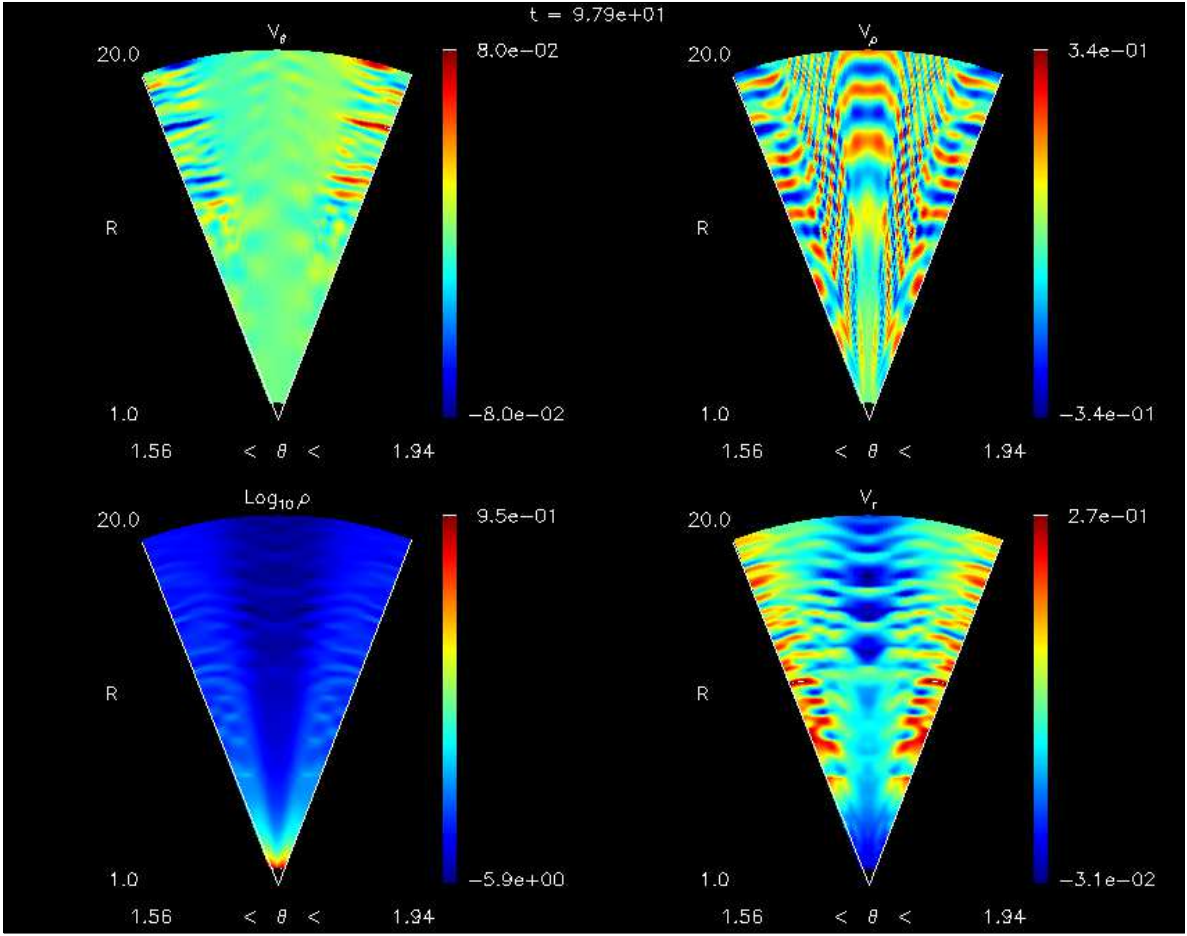,width=0.75\linewidth,angle=90}
\end{tabular}
\caption{
Variation of the radial and azimuthal velocities, $V_\theta$,
$V_\phi$, $V_r$ and $\rho$ in a $r-\theta$ slice of a magnetically open
region from 1--20~$R_{\star}$. The Alfv\'en waves are reflected at a height
of $\sim 5.5~R_{\star}$, which thus marks the onset of the wind.
}
\end{figure*}

\clearpage

\begin{table}
\centering 
\caption{Wind Properties of $\alpha$~Tau (K5~III), $\alpha$~Ori (M2~Iab) and 31~Cyg (K4~Ib)}
\begin{tabular}{c c c c c} 
\noalign{\smallskip}\hline\noalign{\smallskip}
Star & $R_{\star}/R_{\odot}$ & $V_{\rm esc}$ & $V_{\infty}$ & $-$$\dot{M}$ ($M_{\odot}$~yr$^{-1}$) \\ [1.5ex]
\noalign{\smallskip}\hline\noalign{\smallskip}
$\alpha$ Tau & 44  & 115 km~s$^{-1}$ & 30  km~s$^{-1}$  & 1.6 $\times$ 10$^{-11}$ \\
$\alpha$ Ori & 955 &  64 km~s$^{-1}$ & 10  km~s$^{-1}$  & 2 $\times$ 10$^{-6}$ \\
 31 Cyg      & 197 & 152 km~s$^{-1}$ & 90  km~s$^{-1}$  & 3 $\times$ 10$^{-8}$ \\
\noalign{\smallskip}\hline\noalign{\smallskip}
\end{tabular}
\end{table}

\end{document}